\documentclass[a4paper,fleqn]{cas-dc}

\usepackage[numbers,sort&compress]{natbib}
\usepackage{siunitx}
\usepackage{pifont,amssymb}
\usepackage{multirow}
\usepackage{color}
\usepackage{amsmath}
\usepackage{caption}
\usepackage[switch]{lineno}
\usepackage{microtype}
\usepackage{setspace}
\def\tsc#1{\csdef{#1}{\textsc{\lowercase{#1}}\xspace}}
\tsc{WGM}
\tsc{QE}

\begin{document}
\let\WriteBookmarks\relax
\def\floatpagepagefraction{1}
\def\textpagefraction{.001}

\shorttitle{}

\shortauthors{Jinsong Zhang et al.}  

\title [mode = title]{Comparative Analysis of Non-Newtonian Effects on Temporal and Spatial Characteristics of Droplet Generation: Non-Newtonian Fluid as Dispersed or Continuous Phase in Coaxial Two-Phase Flow}  

\author[1]{Jinsong Zhang}
            
\author[1]{Hanhua Song}

\author[2]{Z. L. Wang}[orcid=0000-0002-8382-4800]
\ead{wng_zh@i.shu.edu.cn}
\cormark[1]

\affiliation[1]{organization={School of Mechatronic Engineering and Automation, Shanghai University},
            addressline={No.149 Yanchang Road}, 
            city={Shanghai},
            postcode={200072}, 
            country={P. R. China}}

\affiliation[2]{organization={Shanghai Key Laboratory of Mechanics in Energy Engineering, Shanghai Institute of Applied Mathematics and Mechanics, School of Mechanics and Engineering Science, Shanghai University},           
            city={Shanghai},
            postcode={200444}, 
            country={P. R. China}}

\cortext[1]{Corresponding author}

\begin{abstract}
Comparative Analysis on temporal and spatial behaviors of droplets produced in a converging co-flow has been investigated when  interchanging of phases, NaAlg (non-Newtonian) and soybean oil (Newtonian). The Carreau model is promoted and gives rarely reported negative non-Newtonian index, $n<0$, by which phase diagrams of "butterfly distribution" on temporal $f \cdot \tau \sim\left(Q_d / Q_c\right)^n$ space and "grape distribution" on spatial $d^* / D_c \sim\left(Q_d / Q_c\right)^n$ space are distinguished for the first time. These flow charts show symmetry on refined expression $\left(Q_d / Q_c\right)^n=1$, (either $Q_d / Q_c=1$ or $n=0$) for both comparative experiments. We also find an interesting synchronous transition phenomenon exist, where the interchanging of disperse and continuous phases will not affect their temporal and spatial characteristics of drop generating, which is dynamically rarely happened. 
\end{abstract}

\begin{graphicalabstract}
\includegraphics[scale=0.3]{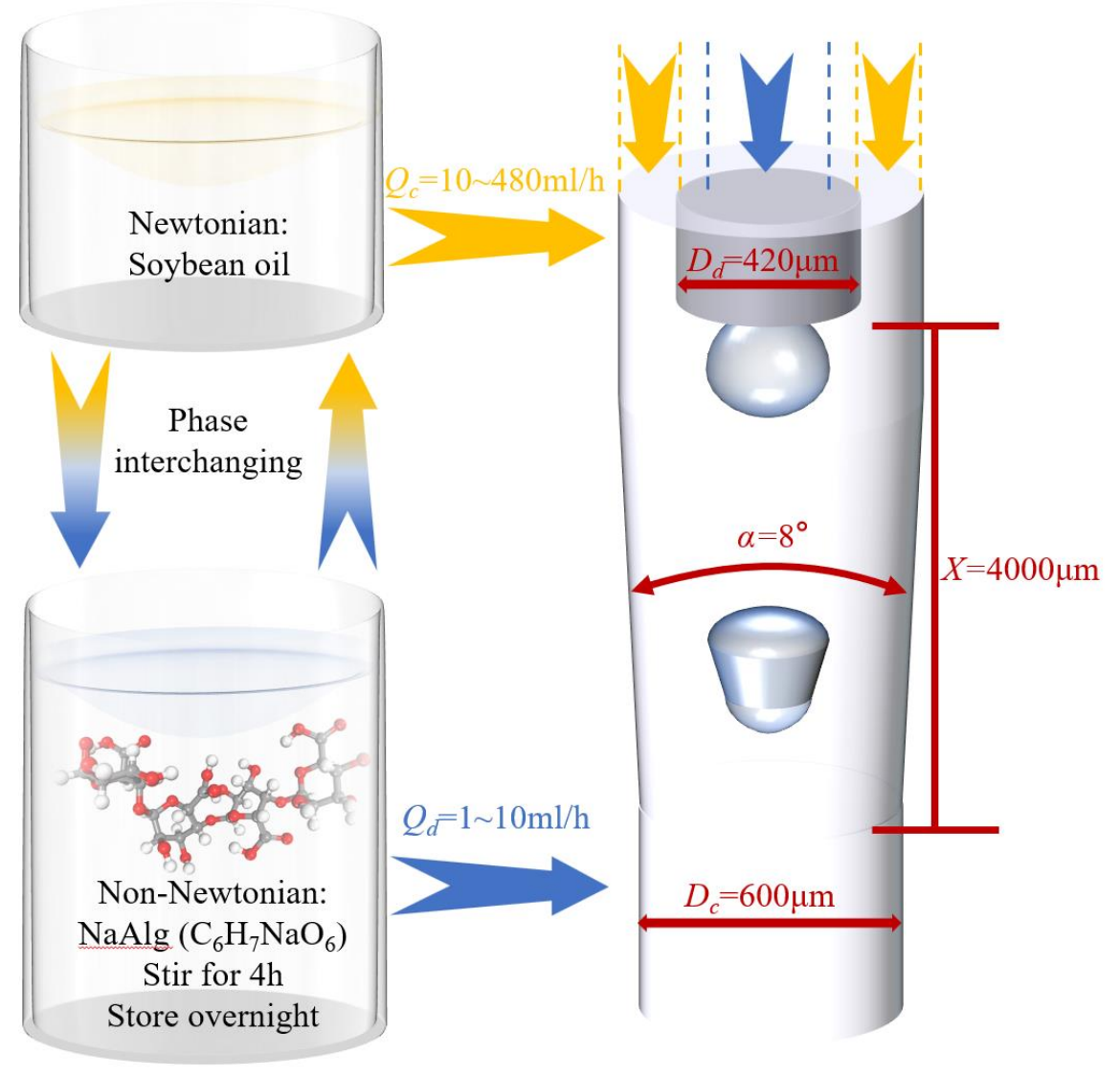}
\includegraphics[scale=0.3]{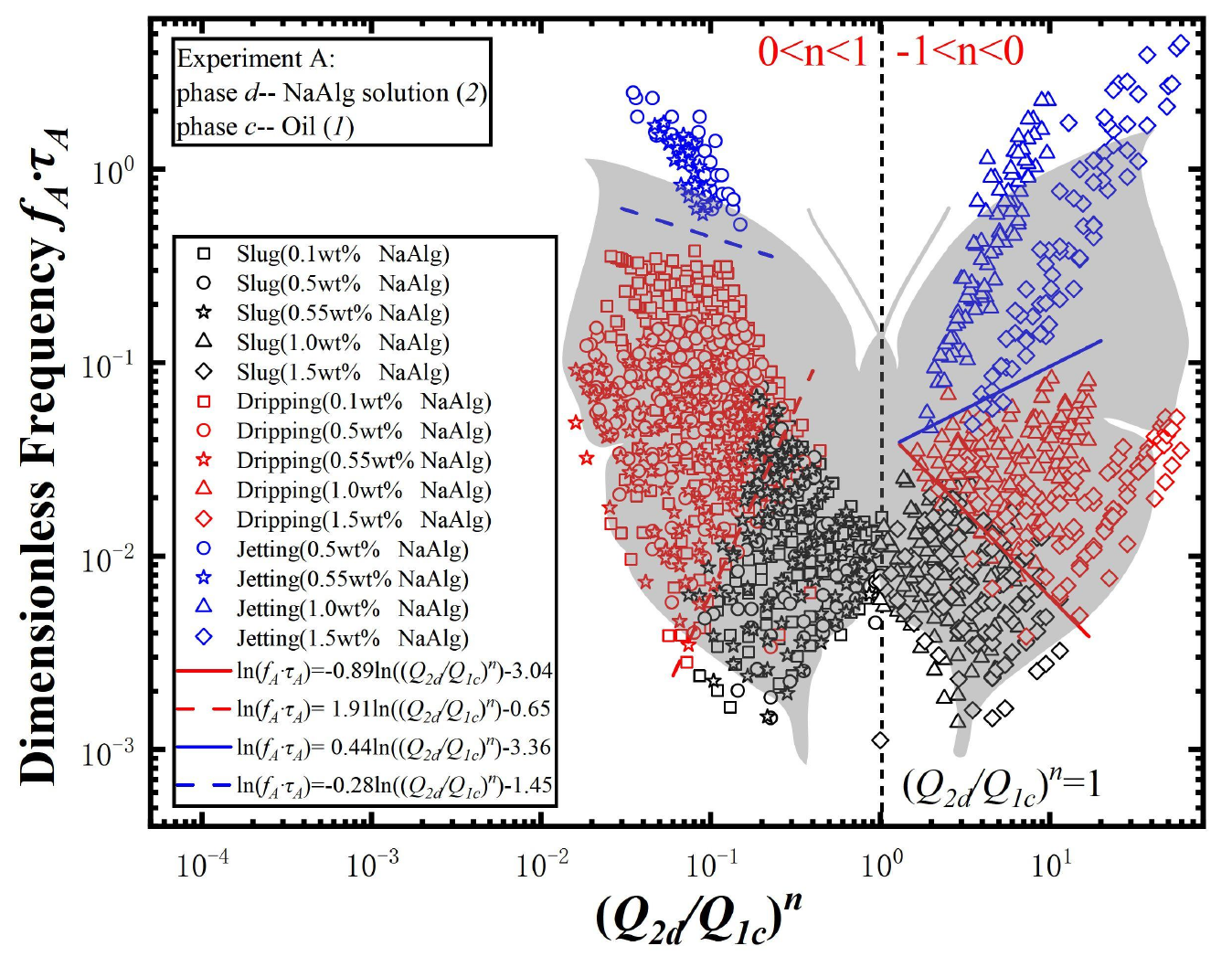}
\end{graphicalabstract}

\begin{highlights}
\item Investigated temporal$\&$spatial behaviors when interchanging of two phases.
\item Modified Carreau model to find non-Newtonian index $n<0$, rarely reported, exist.
\item Found \textit{Butterfly} and \textit{Grape} distributions of separable phase charts, only when introducing of $n$.
\item Discovered temporal$\&$spatial synchronous transition phenomenon.
\end{highlights}


\maketitle

\section{Introduction}\label{sec1}

\begin{table*}[width=0.9\textwidth,htbp]
\begin{tabular*}{\tblwidth}{@{}LLLL@{}}
\hline
\multicolumn{4}{c}{Nomenclature}                                                                                                 \\ \hline
$D$             & Channel diameter (\SI{}{\mu m})            & $\eta$         & Apparent viscosity (\SI{}{Pa\cdot s})            \\
$n$             & Non-Newtonian index                        & $\eta _0$      & Zero-shear-rate viscosity (\SI{}{Pa\cdot s})     \\
$\lambda$       & Material relaxation time (\SI{}{s})        & $\eta _\infty$ & Infinity-shear-rate viscosity (\SI{}{Pa\cdot s}) \\ \cline{3-4}
$Q$             & Flow rate (\SI{}{ml/h})                    & \multicolumn{2}{c}{Dimensionless numbers}                         \\ \cline{3-4}
$u$             & Velocity (\SI{}{m/s})                      & $Ca$           & Capillary number                                 \\
$f$             & Droplet frequency (\SI{}{s^{-1}})          & $We$           & Weber number                                     \\
$d^*$           & Droplet equivalent diameter (\SI{}{\mu m}) & $Re$           & Reynolds number                                  \\ \hline
\multicolumn{2}{c}{Greek symbols}                            & \multicolumn{2}{c}{Subscript}                                     \\ \hline
$\rho$          & Density (\SI{}{kg/m^3})                    & $1$            & Newtonian fluid                                  \\
$\sigma$        & Interfacial tension (\SI{}{N/m})           & $2$            & Non-Newtonian fluid                              \\
$\tau$          & Capillary time (\SI{}{s})                  & $d$            & Dispersed phase                                  \\
$\dot{\gamma }$ & Shear rate (\SI{}{s^{-1}})                 & $c$            & Continuous phase                                 \\ \hline
\end{tabular*}
\end{table*}

Microfluidics \cite{Lari2021mic,Su2021mic,Tan2021go} pertains to the application of scientific and technological principles to control and manipulate micro digital fluids—ranging from picoliters to nanoliters—via microscale channels, spanning from immunoassay \cite{Huang2018an} and bioanalysis \cite{Zheng2003scr} to fluidic optics \cite{Hung2006alt}, and beyond.

A crucial research subfield within microfluidics is liquid-liquid two-phase flow \cite{Zhai2020det,Qian2019a}. Techniques using two-phase flow for droplet generation in microchannels are widely used in diverse arenas, such as drug delivery \cite{Xu2009pre}, cell encapsulation \cite{Zhao2009gen}, protein crystallization \cite{Zheng2004for}, and polymer microcapsules \cite{Fu2016bre}.

Typically, the microdevices for two-phase flow include cross-flowing devices (such as T-shaped \cite{Kovalev2018flow,Timung2015cap} and Y-shaped \cite{Dang2013for,Yin2018stu} junctions), co-flowing devices \cite{Sontti2018for,Deng2017num} (such as the converging coaxial microchannel in this study \cite{Wang2015spe,Wang2022uni}), and flow focusing devices \cite{Wu2015dra,Chen2017exp} (such as cross-shaped and coaxial junctions). Microfluidic devices are commonly fabricated from materials like polydimethylsiloxane (PDMS), quartz glass, and polymethyl methacrylate (PMMA).

Research on the liquid-liquid two-phase flow of Newtonian fluids is prolific, with the flow pattern providing critical insights into the boundaries between various flow patterns and their generative mechanisms. Experimental parameters (i.e. velocity and flow rate) \cite{Wang2021exp,Lee2021sur} and dimensionless numbers (i.e. Capillary number, Weber number, and Reynolds number) \cite{Zhang2020exp,Verma2020eff} can be used to characterize these flow patterns, consequently delineating the boundaries with clarity. Many studies on the flow and mass-transfer mechanisms of two-phase flow in various microdevices explored the flow, shape, breakup, and coalescence of droplets, covered theoretical, experimental, and numerical approaches, recognized the flow patterns under diverse conditions and scrutinized the factors influencing these patterns \cite{Cerdeira2020rev}.

Non-Newtonian \cite{Lakzian2020num,Picchi2018sta}/Newtonian fluids two-phase flow represents an emerging focus within two-phase flow research fields. While non-Newtonian fluids find extensive applications across industries, including food \cite{Xie2012rhe}, energy \cite{Schneider2020rhe}, and biology \cite{Zhao2011two} sectors, however, their unique properties such as shear-thinning/shear-thickening \cite{Granados2021app}, Weissenberg effect \cite{Huang2019on}, and Tom’s phenomenon \cite{Sokhal2018eff} add complexity to related studies. To comprehend these properties, researchers have deployed rheological parameters like viscosity ratio \cite{Fu2015flow}, material relaxation time \cite{Agarwal2020dyn}, density ratio \cite{Dziubinski2004the}, and linked microchannel dimensions, dimensionless numbers, flow rate ratio with experimental phenomena such as droplet size, frequency, and flow patterns. Fu et al. \cite{Fu2015flow} explored the flow patterns for cyclohexane-CMC two-phase flow in T-shaped rectangular microchannels, found that slug, droplet, parallel, and jet flow are the main flow patterns, studied the influence of the aspect ratio of microchannels and the mass fraction of solution on flow patterns, and plotted the  flow pattern diagram. Venu et al. \cite{Agarwal2020dyn} investigated the flow patterns for oil-xanthan solution two-phase flow in a T-shaped microfluidic device, utilized the Carreau-Yasuda model to describe shear-thinning behavior of a non-Newtonian fluid, found that PF, DC, DTJ flow patterns occurred, and included the Carreau number to plot the  flow pattern map. Vagner et al. \cite{Vagner2017for} employed numerical modeling to analyze droplet characteristics of non-Newtonian/Newtonian fluids two-phase flow in coaxial microchannels, found a reduction in calculated droplet size with increased capillary flow speed ratio. Moreover, the droplet size showed a weak dependency on the velocity ratio of the continuous and dispersed phase. Taassob et al. \cite{Taassob2017mon} studied oil-xanthan solution flow patterns in a co-flow device, found that dripping and jetting flow patterns occurred, and examined the impact of dispersed phase viscosity and continuous phase speed on droplet formation. Khater et al. \cite{Khater2020pic} examined the influence of the Capillary number, flow rate ratio, and agar mass fraction on light mineral oil-agar two-phase flow droplet formation in a flow-focusing microfluidic device, and their results suggested a decrease in droplet diameter with increasing agar mass fraction, and highlighted the Capillary number and flow speed ratio as primary factors in droplet volume and formation. Bai et al. \cite{Bai2021gen} conducted a three-dimensional computational study of Janus droplet formation in a double Y-type microfluidic device, identified five different states of tubbing, jetting, intermediate, dripping, and unstable dripping flow pattern under various flow conditions and noted the correlation between inlet flow speed and droplet size under low Capillary number conditions. Battat et al. \cite{Battat2022non} reviewed the generation of bubbles and droplets under varied Capillary number conditions across different microfluidic devices. Despite the diversity and complexity of non-Newtonian fluid two-phase flow, most research concentrates on power-law fluids \cite{Kumar2020inf} rather than Carreau fluids and predominantly uses cross-flow structures in microchannels. The flow pattern boundaries are generally the function of non-Newtonian fluid mass fraction, however, the correlations between flow patterns, droplet temporal and spatial characteristics, and non-Newtonian index $n$, and changes in solution mass fraction are seldom considered. Notably, no comparative studies on phase interchanging experiments have been reported.

Numerous studies, characterized by “front-end stretching and rear-end squeezing” mechanism, on two-phase flow in converging microchannels were conducted \cite{Wang2015spe,Wang2022uni,Zhang2020exp}. The unified temporal and spatial operation formula for all periodic monodisperse droplet (slug, dripping, and jetting) generation presented an outcome in microfluidics, thereby providing superior control capabilities. This study investigates the temporal and spatial characteristics and flow patterns forming by monodisperse microdroplets, focusing on two-phase phase interchanging experiments involving non-Newtonian (NaAlg) and Newtonian (soybean oil) fluids in the convergent coaxial microchannel. Firstly, the Carreau model is modified and the relationship between the non-Newtonian index $n$ and the mass fraction of NaAlg solution is studied. Subsequently, the dimensionless droplet frequency $f \cdot \tau$ and dimensionless equivalent diameter $d^* / D_c$ are introduced, and $f \cdot \tau \sim\left(Q_d / Q_c\right)^n$ and $d^* / D_c \sim\left(Q_d / Q_c\right)^n$ phase diagrams are plotted. Lastly, a synchronous transition phenomenon, where the interchanging of disperse and continuous phases will not affect their temporal and spatial characteristics of drop generating, is identified.

\section{Materials and experimental setups}\label{sec2}

\begin{figure*}[htbp]
	\centering
		\includegraphics[scale=0.7]{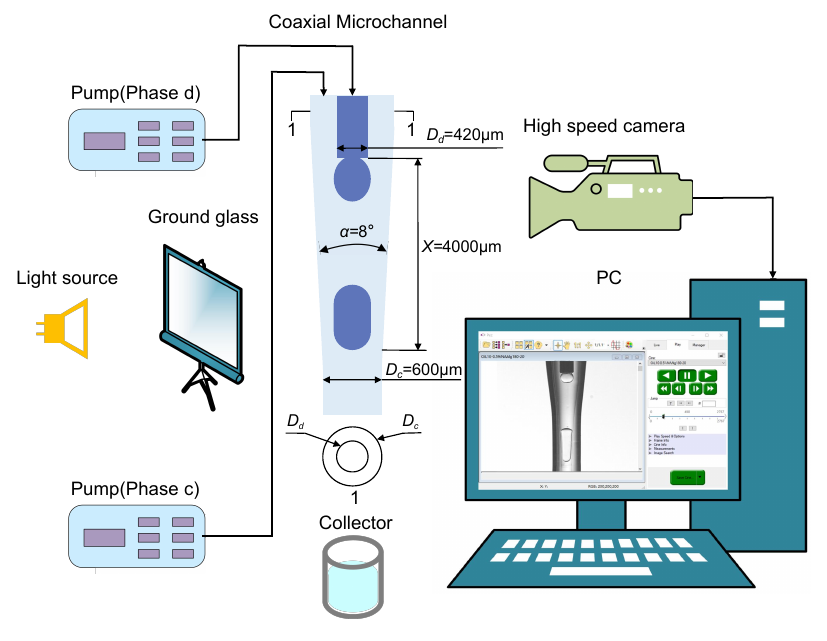}
	  \caption{The schematic microfluidic device.}\label{fig1}
\end{figure*}

The experimental setups for non-Newtonian/Newtonian fluids two-phase flow, shown in Fig. \ref{fig1}, include three main sections: power, microchannel, and observation. The power section includes two peristaltic pumps (AP-0010, SP-6015, Sanotac). The microchannel section incorporates an inner needle (dispersed phase channel), an outer capillary glass tube (continuous phase channel), a PTFE tube, and a collector. The observation section integrates a supplementary light source, ground glass, a high-speed camera (Phantom V611-16G-M, Ametek) with a micro-lens (AT-X M100 PRO D, Tokina), and a computer system. The frame rate for the high-speed camera is \SI{3000}{fps} and the experimental temperature is \SI{25}{\degreeCelsius}.

The size of the microchannel is shown in Fig. \ref{fig1}, the continuous phase channel diameter $D_c=600\mu$m, dispersed phase channel diameter $D_d=420\mu$m, convergence angle $\alpha =$\SI{8}{\degree}, and nozzle injection length $X=4000\mu$m. During the experimentation process, the liquid of the continuous phase (phase $c$) is first injected into the outer capillary glass tube by the peristaltic pump, and this process continues until the liquid saturates the outer capillary glass tube and achieves stability. Subsequently, the liquid of the dispersed phase is injected into the inner needle of the microchannel. Adjust the experimental parameters to achieve a steady observation by holding the dispersed phase flow rate $Q_d$ and changing the continuous phase flow rate $Q_c$, then use the high-speed camera to capture images of the two-phase flow.

Detailed specifications of the experimental design are depicted in Table \ref{tab1}. The soybean oil (CP2015, Jinxiangyaofu) is used as the Newtonian fluids and various mass fractions (\SI{0.1}{wt\%}, \SI{0.5}{wt\%}, \SI{0.55}{wt\%}, \SI{1.0}{wt\%}, \SI{1.5}{wt\%}) of NaAlg solution (S11053, Shyuanye) are used as the non-Newtonian fluid. Two distinctive experiments were established, with an interchanging of the dispersed and continuous phases: In Experiment A, the NaAlg solution serves as the dispersed phase and oil functions as the continuous phase, the flow rate of the dispersed phase $Q_d$ ranges from 1 to \SI{10}{ml/h}, and the flow rate of the continuous phase $Q_c$ from 10 to \SI{480}{ml/h}; In Experiment B, oil is designated as the dispersed phase and the NaAlg solution as the continuous phase, the flow rate of the dispersed phase $Q_d$ ranges still from 1 to \SI{10}{ml/h}, and the flow rate of the continuous phase $Q_c$ from 1 to \SI{250}{ml/h}.

Images of flow patterns are shown in Table \ref{tab1}. Slug, dripping, and jetting flow patterns, forming monodisperse microdroplets, were observed in the experiments. The slug flow pattern droplets are large and piston-like, the droplet vertical length $l$ at the non-convergent downstream area is larger than or nearly the same as the diameter of the continuous phase microchannel $D_c$, and the boundaries make contact with the channel walls leading to blockages. The dripping flow pattern droplets are spherical or ellipsoidal, the droplet vertical length $l$ at the non-convergent downstream area is either less than or approximate to the diameter of the continuous phase microchannel $D_c$. The jetting flow resembles the dripping flow, the droplet vertical length $l$ at the non-convergent downstream area is obviously less than the diameter of the continuous phase microchannel $D_c$, and the dispersed phase fluid elongates at the inlet.

\begin{table*}[width=0.8\textwidth,htbp]
\caption{The experimental design and flow pattern images.}\label{tab1}
\begin{tabular*}{\tblwidth}{@{}LCCCCCC@{}}
\hline
Experiment   & \multicolumn{3}{c}{A}                                                                                                                                & \multicolumn{3}{c}{B}                                                                                                                         \\ \hline
Flow rates   & \multicolumn{3}{c}{\begin{tabular}[c]{@{}c@{}}\SI{10}{ml/h}$\le Q_{c} \le$\SI{480}{ml/h}\\ \SI{1}{ml/h}$\le Q_{d} \le$\SI{10}{ml/h}\end{tabular}}    & \multicolumn{3}{c}{\begin{tabular}[c]{@{}c@{}}1ml/h$\le Q_{c} \le$\SI{250}{ml/h}\\ \SI{1}{ml/h}$\le Q_{d} \le$\SI{10}{ml/h}\end{tabular}}     \\ \hline
Phase $c$    & \multicolumn{3}{c}{\begin{tabular}[c]{@{}c@{}}Oil\\ (Newtonian fluid--$1$)\end{tabular}}                              & \multicolumn{3}{c}{\begin{tabular}[c]{@{}c@{}}NaAlg solution\\ (non-Newtonian fluid--$2$)\end{tabular}}              \\ \hline
Phase $d$    & \multicolumn{3}{c}{\begin{tabular}[c]{@{}c@{}}NaAlg solution\\ (non-Newtonian fluid--$2$)\end{tabular}}               & \multicolumn{3}{c}{\begin{tabular}[c]{@{}c@{}}Oil\\ (Newtonian fluid--$1$)\end{tabular}}                             \\ \hline
Image        & \begin{minipage}[b]{0.2\columnwidth}
		          \centering
		          \raisebox{-.5\height}{\includegraphics[scale=0.5]{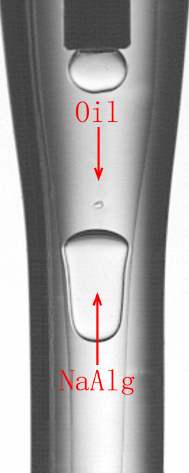}}
	           \end{minipage}
             & \begin{minipage}[b]{0.2\columnwidth}
		          \centering
		          \raisebox{-.5\height}{\includegraphics[scale=0.5]{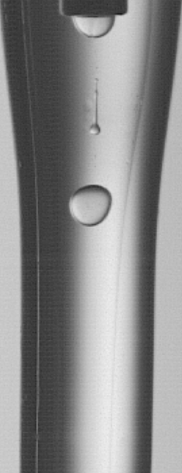}}
	           \end{minipage}
             & \begin{minipage}[b]{0.2\columnwidth}
		          \centering
		          \raisebox{-.5\height}{\includegraphics[scale=0.5]{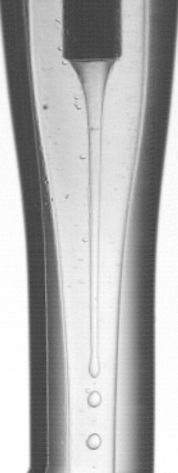}}
	           \end{minipage}
             & \begin{minipage}[b]{0.2\columnwidth}
		          \centering
		          \raisebox{-.5\height}{\includegraphics[scale=0.5]{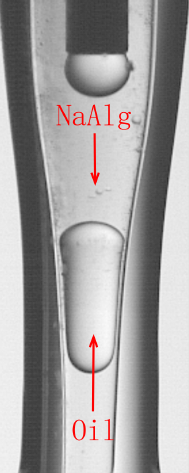}}
	           \end{minipage}
             & \begin{minipage}[b]{0.2\columnwidth}
		          \centering
		          \raisebox{-.5\height}{\includegraphics[scale=0.5]{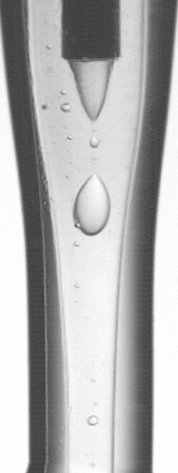}}
	           \end{minipage}
             & \begin{minipage}[b]{0.2\columnwidth}
		          \centering
		          \raisebox{-.5\height}{\includegraphics[scale=0.5]{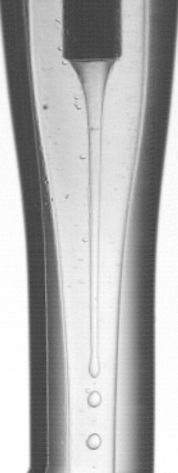}}
	           \end{minipage}                                                                                                                                                                                                                            \\ \hline
Flow pattern &Slug                                 & Dripping                               & Jetting                                & Slug                                 & Dripping                               & Jetting                              \\ \hline
\end{tabular*}
\end{table*}

The detailed material properties are shown in Table \ref{tab2}. The NaAlg solution is prepared as follows: Initially, position a beaker on the electronic balance (JA302, Puchun), add specified quantities of NaAlg powders and deionized water. Subsequently, place the beaker on magnetic stirrer (ZGCJ-3A, Zigui) operating at the speed of \SI{2500}{r/min}. Finally, once the NaAlg powder is fully dissolved, cease the stirring and wait for \SI{24}{h}, ensuring complete degasification. Measure the solution's density. Measure the interfacial tension of the NaAlg solution by tensiometer (CV-ZL1021, Cvok) and the shear viscosity of oil by viscometer (NDJ-5S, Lichen).

Phenomenological constitutive equations used to represent the rheological behavior of the non-Newtonian fluids are the Carreau model \cite{Kokini1984pre}, the Power-Law model \cite{Waele1923vis}, the Cross model \cite{Cross1965rhe}, the Herschel-Bulkley model \cite{Herschel1926kon} and so on. The Carreau model, accounting for both shear-thinning and shear-thickening behaviors, aptly describes the consistent viscosity limits at low shear rates (zero-shear-rate viscosity $\eta_0$) and high shear rates (infinity-shear-rate viscosity $\eta_\infty$) \cite{Picchi2018sta}. The Carreau model is

\begin{equation}\label{equ1}
\begin{aligned}
    \eta=(1+(\lambda\dot{\gamma})^2)^{\frac{n-1}{2}}(\eta_0-\eta_\infty)+\eta_\infty
\end{aligned}
\end{equation}

\noindent where $\eta$ is the apparent viscosity, $\lambda$ is the material relaxation time, $\dot{\gamma}$ is the shear rate and $n$ is the non-Newtonian index. Then the viscosity limits are

\begin{equation}\label{equ2}
\begin{aligned}
    \eta(\dot{\gamma}\to0)\to\eta_0,\eta(\dot{\gamma}\to\infty)\to\eta_\infty
\end{aligned}
\end{equation}

\begin{figure*}[htbp]
	\centering
		\includegraphics[scale=0.5]{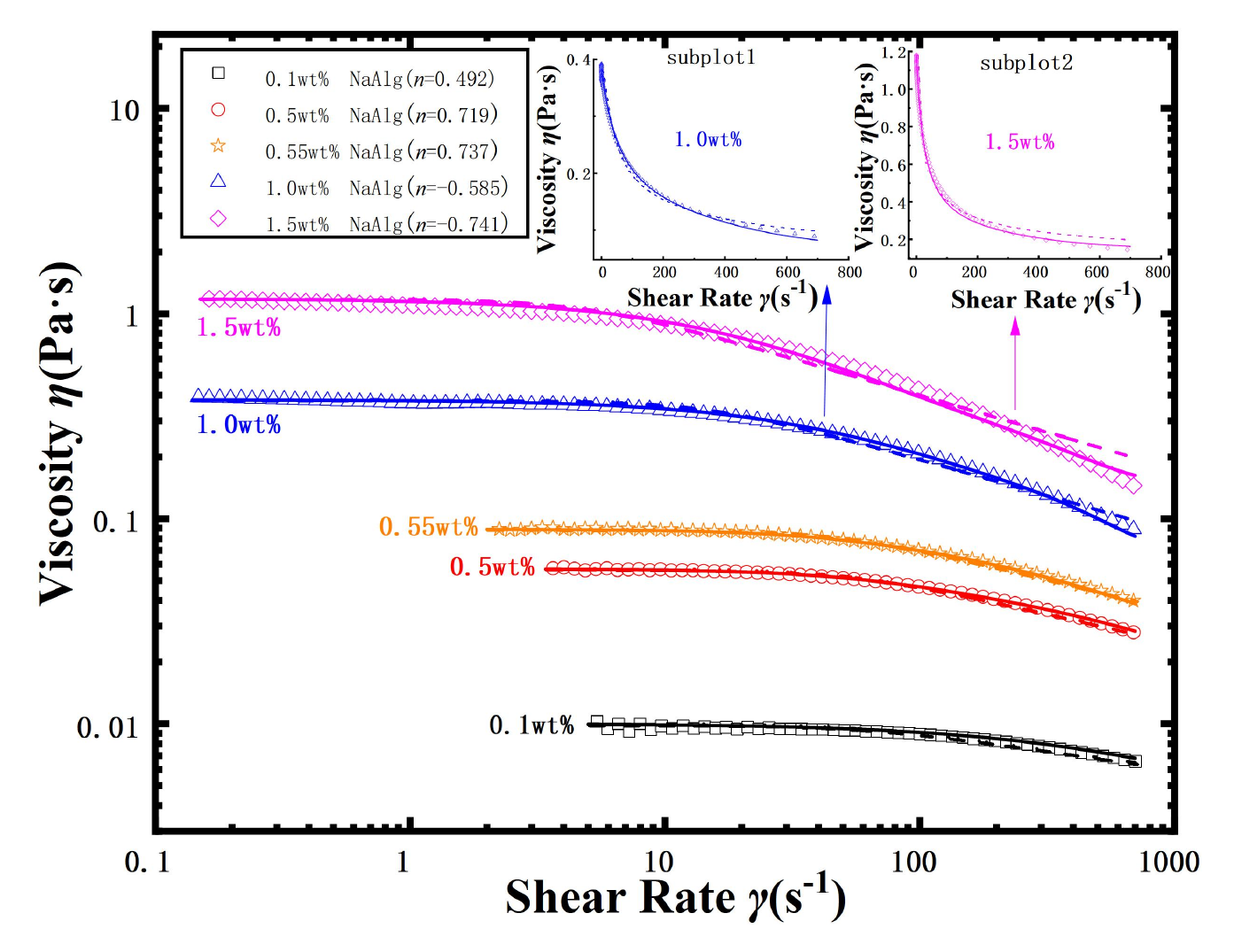}
	  \caption{Rotational rheometer measurement data for NaAlg solutions. NaAlg solution mass fractions used were \SI{0.1}{wt \%}($\square$), \SI{0.5}{wt \%}(\textcolor{red}{$\bigcirc$}), \SI{0.55}{wt \%}(\textcolor{orange}{\ding{73}}), \SI{1.0}{wt \%}(\textcolor{blue}{$\bigtriangleup$}), and \SI{1.5}{wt \%}(\textcolor{magenta}{$\lozenge$}). Characterization results derived from the Carreau model and the modified Carreau model are depicted by dashed and solid lines respectively. Subplot 1 and Subplot 2 offer magnified views of the fitting results for \SI{1.0}{wt \%} and \SI{1.5}{wt \%} NaAlg solutions respectively.}\label{fig2}
\end{figure*}

Rotational rheometer measurement data and the Carreau model fitting results are shown in Fig. \ref{fig2}. The calculation for the Carreau model parameters is as follows: Initially, set the shear rate range of rotational rheometer (MCR92, Anton Paar) from 0.1 to \SI{200}{s^{-1}} and measure the apparent viscosity $\eta$ of the NaAlg solution under various shear rates $\dot{\gamma}$. Secondly, fit the data to a polynomial exponential function, compute the limit values at both zero and infinity shear rates $\dot{\gamma}$, and then estimate the zero-shear-rate viscosity $\eta_0$ and infinity-shear-rate viscosity $\eta_\infty$ for different mass fraction of the NaAlg solution, which are shown in Table \ref{tab2}. Subsequently, use the multiple regression method to calculate the material relaxation time $\lambda$ and non-Newtonian index $n$ by rheological data.

As shown in Fig. \ref{fig2}, the apparent viscosity $\eta$ escalates as the mass fraction increases. The subplots highlight that when the mass fraction is large (\SI{1.0}{wt\%}, \SI{1.5}{wt\%}) and the corresponding viscosity $\eta$ is high, the Carreau model exhibits considerable deviations as the shear rate $\dot{\gamma}$ nears infinity, particularly when the mass fraction is larger, and the deviation is greater. To enhance the fitting accuracy, modify the Carreau model, which can be expressed as

\begin{equation}\label{equ3}
\begin{aligned}
    \eta=(1+\lambda_1\dot{\gamma}+(\lambda_2\dot{\gamma})^2+(\lambda_3\dot{\gamma})^3+(\lambda_4\dot{\gamma})^4)^{\frac{n-1}{2}}(\eta_0-\eta_\infty)+\eta_\infty 
\end{aligned}
\end{equation}

\noindent where $\lambda _{1}$, $\lambda _{2}$, $\lambda _{3}$, $\lambda _{4}$ are the polynomial material relaxation time.

Parameters of the modified Carreau model ($\lambda _{1}$, $\lambda _{2}$, $\lambda _{3}$, $\lambda _{4}$, $n$) are shown in Table \ref{tab2}. Normally, Newtonian fluids are denoted as $n=1$, and shear-thickening non-Newtonian fluids as $n>1$ \cite{Dhiman2019hyd}, shear-thinning non-Newtonian fluids as $0<n<1$ \cite{Airiau2020flo}, with scarce literature documenting $n<0$ \cite{Suresh2016rhe}. As shown in Fig. \ref{fig2}, the apparent viscosity $\eta$ decreases with an increasing shear rate $\dot{\gamma}$, and the non-Newtonian index $n<1$, which indicate that the NaAlg solutions are shear-thinning non-Newtonian fluids, and additionally, the non-Newtonian index $n$ shifts from positive to negative as the mass fraction increases.

\begin{table*}[width=0.9\textwidth,htbp]
\caption{Materials physical properties and Carreau model parameters.}\label{tab2}
\begin{tabular*}{\tblwidth}{@{}LLLLLLLLLL@{}}
\hline
Fluid                 & $\rho$(\SI{}{kg/m^3}) & $\sigma$(\SI{}{N/m}) & $\eta _{0}$(\SI{}{Pa \cdot s}) & $\eta _{\infty}$(\SI{}{Pa \cdot s}) & $\lambda_1 $(\SI{}{s}) & $\lambda_2 $(\SI{}{s}) & $\lambda_3 $(\SI{}{s}) & $\lambda_4 $(\SI{}{s}) & $n$    \\ \hline
Oil                   & 893                   & -                    & 0.063                          & 0.063                               & -                      & -                      & -                      & -                      & 1      \\
\SI{0.1}{wt \%}NaAlg  & 989                   & 0.0223               & 0.010                          & 0.004                               & 0.008                  & 0.003                  & 0.003                  & 0                      & 0.492  \\
\SI{0.5}{wt \%}NaAlg  & 999                   & 0.0231               & 0.057                          & 0.005                               & 0.010                  & 0.017                  & 0                      & 0.005                  & 0.719  \\
\SI{0.55}{wt \%}NaAlg & 1001                  & 0.0234               & 0.089                          & 0.006                               & 0.011                  & 0.022                  & 0                      & 0.008                  & 0.737  \\
\SI{1.0}{wt \%}NaAlg  & 1007                  & 0.0245               & 0.380                          & 0.062                               & 0.017                  & 0                      & 0                      & 0.003                  & -0.585 \\
\SI{1.5}{wt \%}NaAlg  & 1017                  & 0.0250               & 1.181                          & 0.142                               & 0.041                  & 0                      & 0                      & 0.004                  & -0.741 \\ \hline
\end{tabular*}
\end{table*}

\section{Results and discussion}\label{sec3}
\subsection{Limitations of traditional methods on phase diagrams for temporal and spatial behaviors}\label{sec3.1}

The frequency and size of droplets are the key parameters to study the droplet temporal and spatial behaviors. Dimensionless droplet frequency $f \cdot \tau$ is frequently used by scholars to study the temporal behaviors of droplets \cite{Liu2018for}. The instances of the first $t_1$ and $i$-th $t_i$ droplets passing through a fixed position in the microchannel are recorded to calculate the droplet frequency $f=\frac{i-1}{t_i-t_1}$. The capillary time $\tau=\sqrt{\rho_cD_c^3/\sigma}$ refers to the time required for surface tension to influence dispersed phase droplets \cite{Du2018bre}. Given that the dimension of $f$ is the inverse of time $(T^{-1})$, multiply it by $\tau(T)$ for non-dimensionalization. Consequently, the definition of the dimensionless droplet frequency is

\begin{equation}\label{equ4}
\begin{aligned}
    f\cdot\tau=f\cdot\sqrt{\frac{\rho_cD_c^3}\sigma}
\end{aligned}
\end{equation}

The convergent structure of the microchannel results in the inaccuracies in measuring the length of droplets. Consequently, droplets are modeled as spheres of equivalent volume, so that the droplet volume $V=Q_d/f$ can be represented by the sphere's radius $r$ as $V=4\pi r^3/3$. The sphere's diameter is regarded as the dispersed phase equivalent droplet diameter $d^*=2\sqrt[3]{3Q_d/4\pi f}$. $d^*$ is divided by the continuous phase microchannel diameter $D_c$ to be dimensionless, so the droplet dimensionless equivalent diameter is defined as

\begin{equation}\label{equ5}
\begin{aligned}
    d^*/D_c=\sqrt[3]{\frac{6Q_d}{\pi f}}/D_c
\end{aligned}
\end{equation}

\begin{figure*}[htbp]
	\centering
		\includegraphics[scale=0.7]{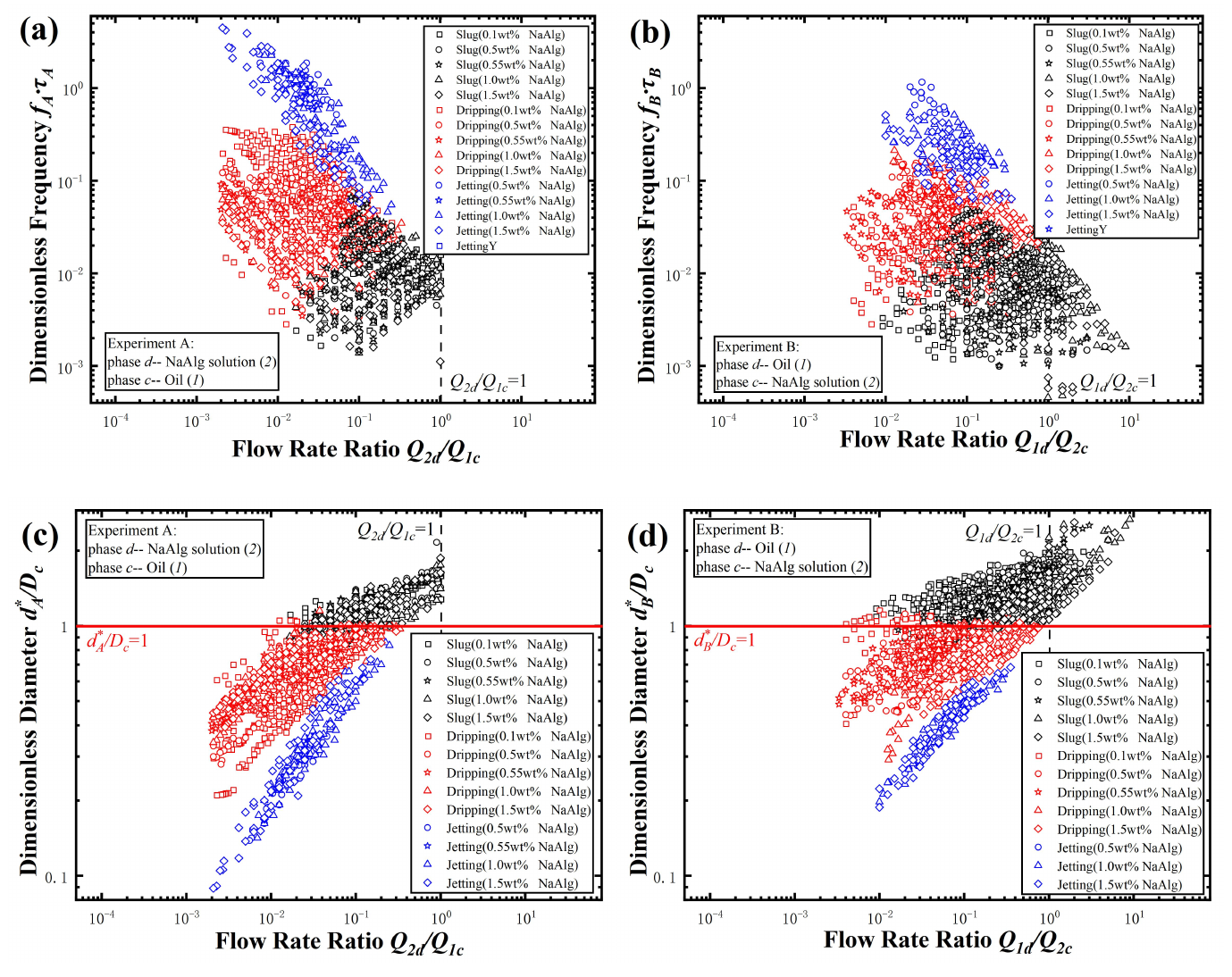}
	  \caption{$f \cdot \tau \sim Q_d / Q_c$ and $d^* / D_c \sim Q_d / Q_c$ phase diagrams for different mass fractions NaAlg solutions in two-phase interchanging experiments. NaAlg solution mass fractions used were \SI{0.1}{wt \%}($\square$), \SI{0.5}{wt \%}($\bigcirc$), \SI{0.55}{wt \%}(\ding{73}), \SI{1.0}{wt \%}($\bigtriangleup$), and \SI{1.5}{wt \%}($\lozenge$). Slug (black), dripping (red), and jetting (blue) flow patterns were observed in the experiments. Dashed lines represent $Q_d / Q_c=1(Q_d=Q_c)$, and red solid lines represent $d^* / D_c=1$, which is the boundary between slug and other flow regimes. (a) $f_{A}  \cdot \tau_{A} \sim Q_{2d} / Q_{1c}$ phase diagram. In Experiment A, NaAlg solution serves as the dispersed phase and oil as the continuous phase. (b) $f_{B}  \cdot \tau_{B} \sim Q_{1d} / Q_{2c}$ phase diagram. In Experiment B, oil serves as the dispersed phase and NaAlg solution as the continuous phase. (c) $d_{A}^{*}  / D_c \sim Q_{2d} / Q_{1c}$ phase diagram in Experiment A. (d) $d_{B}^{*}  / D_c \sim Q_{1d} / Q_{2c}$ phase diagram in Experiment B.}\label{fig3}
\end{figure*}

Newtonian/Newtonian fluids two-phase flow commonly uses parameters such as velocity, flow rate, or dimensionless numbers to delineate flow regimes. Wang et al. \cite{Wang2022uni} established the convergence angle $\alpha$ and needle displacement $x$, created diagrams illustrating four various flow regimes (slug, dripping, thin jet, saugage) at different dispersed phase flow rates $Q_d$. Discovered that the boundaries shift in response to changes in $x$, using the continuous phase Capillary number $Ca_c$ and the dispersed phase Weber number $We_d$ to analyze the transition between Pan-dripping and Pan-jetting regions. Nevertheless, the inherent complexities of non-Newtonian fluids render conventional analysis methods more challenging. Bai et al. \cite{Bai2021gen} plotted flow regime maps with the Capillary numbers of both phases $Ca_c$ and $Ca_d$. Fu et al. \cite{Fu2015flow} used the superficial velocity of the continuous $u_c$ and dispersed phases $u_d$ to create flow regime maps, revealing that changes in the mass fraction of CMC aqueous solutions influence the boundaries of flow regimes. The frequency and diameter of the droplets are often non-dimensionalized to study the relationship among droplet frequency, size, and flow regime. Taassob et al. \cite{Taassob2017mon} explored the relationship between the droplet dimensionless radius and $u_c/u_d$, uncovering a jumping transition zone between dripping to jetting regimes. As shown in Fig. \ref{fig3}, the relationship between droplet frequency, droplet size, flow rate, and flow regime is studied by drawing $f \cdot \tau \sim Q_d / Q_c$ and $d^* / D_c \sim Q_d / Q_c$ phase diagrams for different mass fractions of NaAlg solutions based on conventional research methods.

As shown in Fig. \ref{fig3}, the study becomes complex after introducing $f \cdot \tau$ and $d^* / D_c$ because of the fluctuates of mass fractions. In order to tackle this challenge, key parameters of non-Newtonian fluids were introduced, including viscosity ratio $\eta_c/\eta_d$, material relaxation time $\lambda$, non-Newtonian index $n$, density ratio $\rho_c/\rho_d$ and so on. Consequently, the significant influences of $n$ were realized, so that $n$ was integrated into the study to further examine the temporal and spatial behaviors of droplets.

\subsection{Significance of the non-Newtonian index $n$ in distributing temporal and spatial behaviors}\label{sec3.2}

\subsubsection{"Butterfly" in $f \cdot \tau \sim\left(Q_d / Q_c\right)^n$ phase diagrams}\label{sec3.2.1}

\begin{figure*}[htbp]
	\centering
		\includegraphics[scale=0.6]{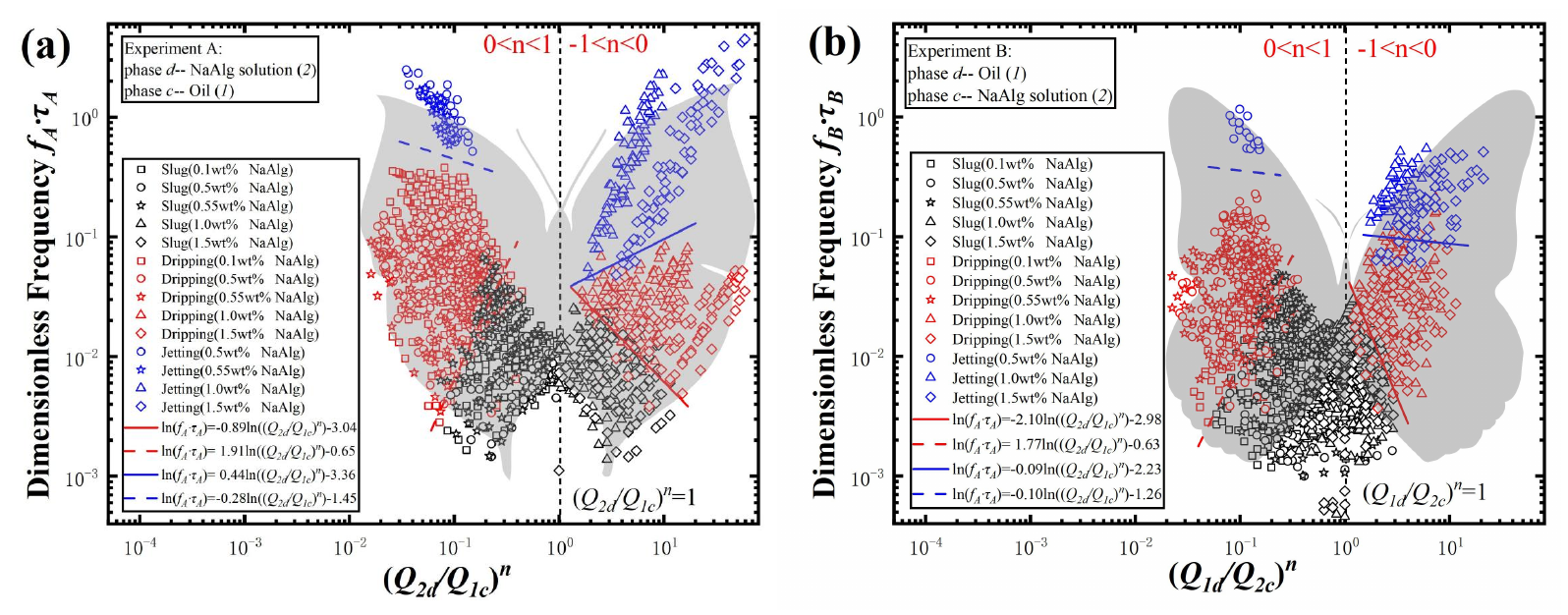}
	  \caption{$f \cdot \tau \sim\left(Q_d / Q_c\right)^n$ phase diagrams for NaAlg solutions of different mass fractions in two-phase interchanging experiments. NaAlg solution mass fractions used were \SI{0.1}{wt \%}($\square$), \SI{0.5}{wt \%}($\bigcirc$), \SI{0.55}{wt \%}(\ding{73}), \SI{1.0}{wt \%}($\bigtriangleup$), and \SI{1.5}{wt \%}($\lozenge$). Slug (black), dripping (red), and jetting (blue) flow patterns were observed in the experiments. Dashed lines represent $\left(Q_d / Q_c\right)^n=1$, for either $Q_d / Q_c=1$ or $n=0$, as the symmetry line for the flow regime distribution. The blue line demarcates the transition from jetting to dripping flow, whereas the red line defines the boundary between dripping and slug flow, which are calculated by machine learning of all the data using MATLAB (fitcsvm; Support Vector Machines Function). (a) $f_A \cdot \tau_A \sim\left(Q_{2d} / Q_{1c}\right)^n$ phase diagrams. In Experiment A, NaAlg solution serves as the dispersed phase and oil as the continuous phase. (b) $f_B \cdot \tau_B \sim\left(Q_{1d} / Q_{2c}\right)^n$ phase diagrams. In Experiment B, oil serves as the dispersed phase and NaAlg solution as the continuous phase.}\label{fig4}
\end{figure*}

In order to study the temporal behaviors of droplets and the distribution of flow regimes more effectively, the non-Newtonian index $n$ is introduced and $f \cdot \tau \sim\left(Q_d / Q_c\right)^n$ phase diagrams for NaAlg solutions of different mass fractions is plotted, as shown in Fig. \ref{fig4}.

Some mass fractions of the NaAlg solution fail to yield jetting flow. In Experiment A, the slug flow has the lowest $f$ (metaphorically represented as the butterfly's wing tails), the dripping flow shows slightly higher $f$ than slug flow, while the jetting flow has the highest $f$ (symbolized as the butterfly's wingtips). In Experiment B, the distribution pattern mirrors that in Experiment A. Compared to Experiment A, more slug flows are generated in Experiment B, implying that the production of oil droplets is more easier than that of NaAlg droplets, when the flow rates of both phases remain unchanged ($Q_{cB}=Q_{cA},Q_{dB}=Q_{dA}$).

The range of $(Q_d / Q_c)^n$ in Experiment A is broader than that in Experiment B (leading to larger butterfly wings), as shown in Table \ref{tab3}. When $Q_d/Q_c$ is small enough, the jetting flow is produced. This phenomenon indicates that a decline in apparent viscosity $\eta$ coincides with an increase in shear rate $\dot{\gamma}$ as the velocity of NaAlg solutions $u$ escalates. The combined viscous force and surface tension of the oil sufficiently counteract the inertial force of the NaAlg solution, preventing the continuous phase from shearing the dispersed phase and, consequently, inhibiting large droplet formation.

The range of NaAlg droplets frequency $f_A$ surpasses the range of oil droplets frequency $f_B$ (indicating a higher butterfly position), as shown in Table \ref{tab3}. Considering the capillary time in both experiments are approximate ($\tau_A = \tau_B$), the prominent difference in $f_{A}  \cdot \tau_{A}$ and $f_{B}  \cdot \tau_{B}$ is primarily attributed to the substantial deviation in $f_A$ and $f_B$, and when $f \cdot \tau$ exhibits a larger value, it indicates that the NaAlg solution facilitates droplet formation easier.

The boundaries among slug, dripping, and jetting flow are clear, as shown in Fig. \ref{fig4}. The non-Newtonian index $n$ displays both positive and negative values, causing the NaAlg solution to distribute the flow regimes on both sides of the symmetry axis $\left(Q_d / Q_c\right)^n=1$ (represented as the middle of butterfly's body). As $n$ converges to zero, $\left(Q_d / Q_c\right)^n=1$, for either $Q_d / Q_c=1$ or $n=0$, the flow regime distribution moves closer to the symmetry axis. However, in the experiments conducted, no non-Newtonian index with values of $n=0$ exhibited.

The study reveals that as $Q_d / Q_c$ escalates, $f \cdot \tau$ diminishes, indicating a flow regime shift favoring the genesis of large-scale droplets, which aligns with the results reported by Sontti et al. \cite{Liu2018for}. As $n$ possesses both positive and negative values, the changing patterns in $f \cdot \tau$ and flow regimes are different, as shown in Table \ref{tab3}. Moreover, alterations in the solution mass fraction induce changes in $n$. Sontti et al. \cite{Sontti2017cfd} and Chen et al. \cite{Chen2020mod} discerned that as power-law index of power-law fluids augments, there was a concurrent increase in $\eta$ and $f$, and a reduction in droplet volume, which subtly mirrored in the $f \cdot \tau \sim\left(Q_d / Q_c\right)^n$ phase diagrams, but more prominent in $d^* / D_c \sim\left(Q_d / Q_c\right)^n$ phase diagrams in section \ref{sec3.2.2}. In addition, Liu et al. \cite{Liu2018for} and Chen et al. \cite{Chen2020mod} proposed the use of $Ca_c$ to establish a scaling law between $f$ and $Q_d/Q_c$, proffering a promising avenue for future investigation in this study.

\subsubsection{"Grape" in $d^* / D_c \sim\left(Q_d / Q_c\right)^n$ phase diagrams}\label{sec3.2.2}

\begin{figure*}[htbp]
	\centering
		\includegraphics[scale=0.5]{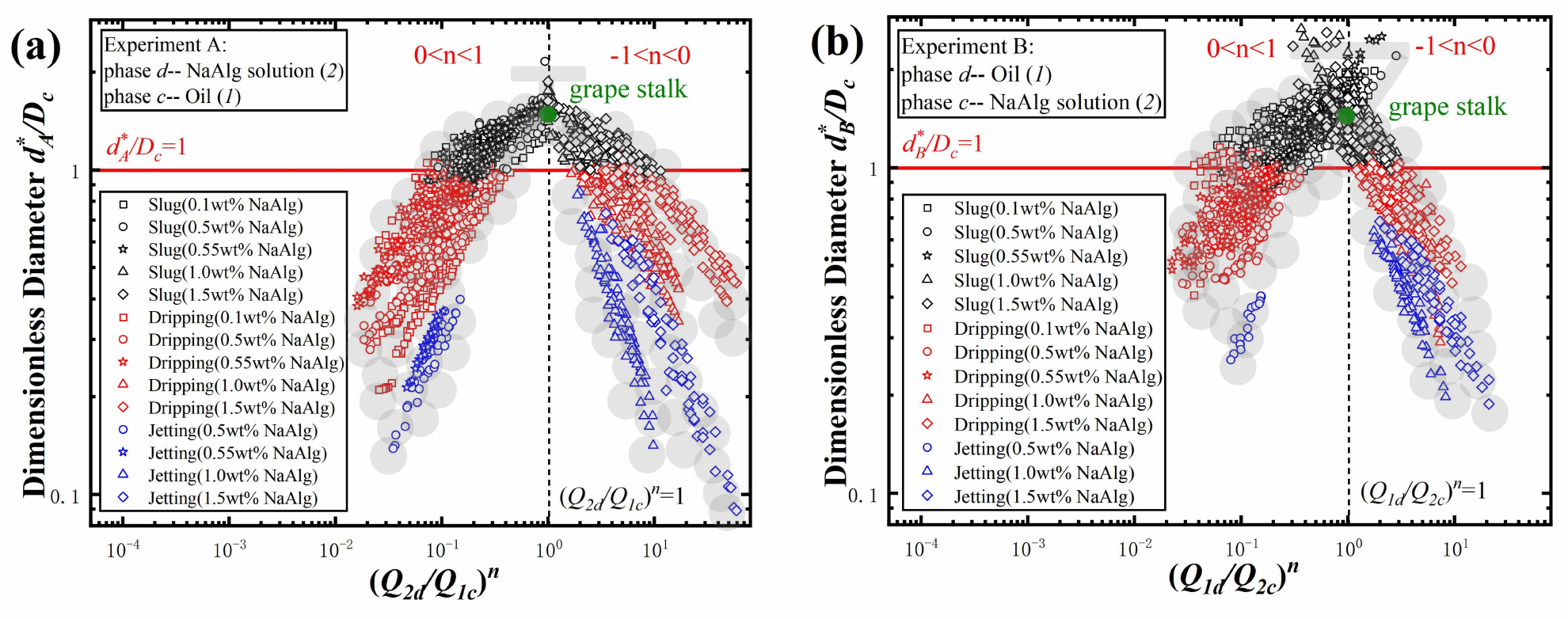}
	  \caption{$d^* / D_c \sim\left(Q_d / Q_c\right)^n$ phase diagrams for NaAlg solutions of different mass fractions in two-phase interchanging experiments. NaAlg solution mass fractions used were \SI{0.1}{wt \%}($\square$), \SI{0.5}{wt \%}($\bigcirc$), \SI{0.55}{wt \%}(\ding{73}), \SI{1.0}{wt \%}($\bigtriangleup$), and \SI{1.5}{wt \%}($\lozenge$). Slug (black), dripping (red), and jetting (blue) flow patterns were observed in the experiments. Dashed lines represent $\left(Q_d / Q_c\right)^n=1$, for either $Q_d / Q_c=1$ or $n=0$, as the symmetry line for the flow regime distribution, and red solid lines represent $d^* / D_c=1$, which is the boundary between slug and other flow regimes. The data converging at a point on the symmetry axis as grape stalk. (a) $d^*_A / D_c \sim\left(Q_{2d} / Q_{1c}\right)^n$ phase diagrams. In Experiment A, NaAlg solution serves as the dispersed phase and oil as the continuous phase. (b) $d^*_B / D_c \sim\left(Q_{1d} / Q_{2c}\right)^n$ phase diagrams. In Experiment B, oil serves as the dispersed phase and NaAlg solution as the continuous phase.}\label{fig5}
\end{figure*}

In order to study the spatial behaviors of droplets more effectively, the non-Newtonian index $n$ is introduced, and $d^* / D_c \sim\left(Q_d / Q_c\right)^n$ phase diagrams for NaAlg solutions of different mass fractions is plotted, as shown in Fig. \ref{fig5}.

The slug flow, distinguishable by its larger $d^*$ (represented at the upper section of the grape bunch), as shown in Fig. \ref{fig5}. On the other hand, the dripping flow exhibits marginally smaller $d^*$, while the jetting flow features the smallest $d^*$ (denoted at the tail of the grape bunch). Experiment B maintains similar distribution patterns to those of Experiment A. There is a clear boundary $d^* / D_c=1$ between slug and dripping flow (dividing the grape bunch into two distinct sections), with the slug flow above, and the dripping and jetting flow below.

The range of NaAlg droplets equivalent diameter $d_A^*$ is narrower than that of oil droplets $d_B^*$ (suggesting a lower grape bunch position), as shown in Table \ref{tab3}. Concurrently, a higher frequency of NaAlg droplet $f_A$ formation is observed in Experiment A. This relationship is expounded by Equation (\ref{equ5}), which establishes an inverse proportionality between $f$ and $d^* / D_c$.

The convergence point on the symmetry axis $\left(Q_d / Q_c\right)^n=1$ (depicted as the grape stalk) is shown in Fig. \ref{fig5}. In Experiment B, particular flow regimes emerge above the convergence point (forming the root of the grape bunch), when $Q_{d}$ surpasses $Q_{c}$. This suggests that the inertial force of the oil struggles to the counterbalance viscous force of NaAlg solution, resulting in oil undergoing shear from the high-viscosity NaAlg solution and generating droplets. Notably, this phenomenon is absent in Experiment A, indicating the increased difficulty of shearing a high-viscosity NaAlg solution when oil is the continuous phase.

Experimental data indicate an augmentation in $d^* / D_c$ with an increasing $Q_d / Q_c$, corroborating the results of previous studies by Fu et al. \cite{Fu2016bre}, Vagner et al. \cite{Vagner2017for}, Khater et al. \cite{Khater2020pic}, and Bai et al. \cite{Bai2021gen}. This trend is attributable to the interplay between the shear stress imposed by the continuous phase fluid and the surface tension of the dispersed phase fluid, wherein an increase in $Q_c$ leads to a higher pressure and velocity gradients at the surface, which results in augmented resistance, prompting the dispersed phase fluid to break down into smaller droplets. Conversely, a surge in $Q_d$ necessitates a greater energy accumulation by the continuous phase to fragment the dispersed phase, thereby facilitating the generation of larger droplets. Considering $n$ can possess both positive and negative values, the changing patterns in $d^* / D_c$ and flow regime are different, as shown in Table \ref{tab3}. Furthermore, Fu et al. \cite{Fu2016bre} and Liu et al. \cite{Liu2018for} introduced $Ca_c$ to establish a scaling law between the dimensionless droplet size and $Q_d / Q_c$, presenting a prospective direction for subsequent research.

\begin{table*}[width=\textwidth,htbp]
\caption{The phenomena and patterns of the “butterfly distribution” and “grape distribution”.}\label{tab3}
\begin{tabular*}{\tblwidth}{@{}LLL@{}}
\hline
Phenomena                                            & Experiment A                                                                                                            & Experiment B                                                                                                        \\ \hline
$\left(Q_d / Q_c\right)^n$                           & $0.01 \sim 60$                                                                                                          & $0.02 \sim 20$                                                                                                      \\
$f \cdot \tau$                                       & $1\times 10^{-3} \sim 4.5$                                                                                                & $5\times 10^{-4} \sim 1.2$                                                                                         \\
$d^* / D_c$                                          & $0.08 \sim 2.2$                                                                                                          & $0.18 \sim 2.7$                                                                                                      \\ \hline
\multirow{2}{*}{Flow pattern distribution}           & \multicolumn{2}{c}{When $0<n<1$, the distribution of flow is confined within $\left(Q_d / Q_c\right)^n<1$.}                                                                                                                                   \\
                                                     & \multicolumn{2}{c}{When $-1<n<0$, the distribution of flow is confined within $\left(Q_d / Q_c\right)^n>1$.}                                                                                                                                  \\ \hline
\multirow{2}{*}{$\left(Q_d / Q_c\right)^n$ increase} & \multicolumn{2}{c}{When $0<n<1$, $f \cdot \tau$ reduces, $d^* / D_c$ augments, and dripping and jetting transfer to slug.}                                                                                                                    \\
                                                     & \multicolumn{2}{c}{When $-1<n<0$, $f \cdot \tau$ augments, $d^* / D_c$ reduces, and slug transfers to dripping and jetting.}                                                                                                                  \\ \hline
\end{tabular*}
\end{table*}

\subsection{Synchronous transition phenomenon of the temporal and spatial behaviors in two-phase interchanging experiments}\label{sec3.3}

\begin{figure*}[htbp]
	\centering
		\includegraphics[scale=0.7]{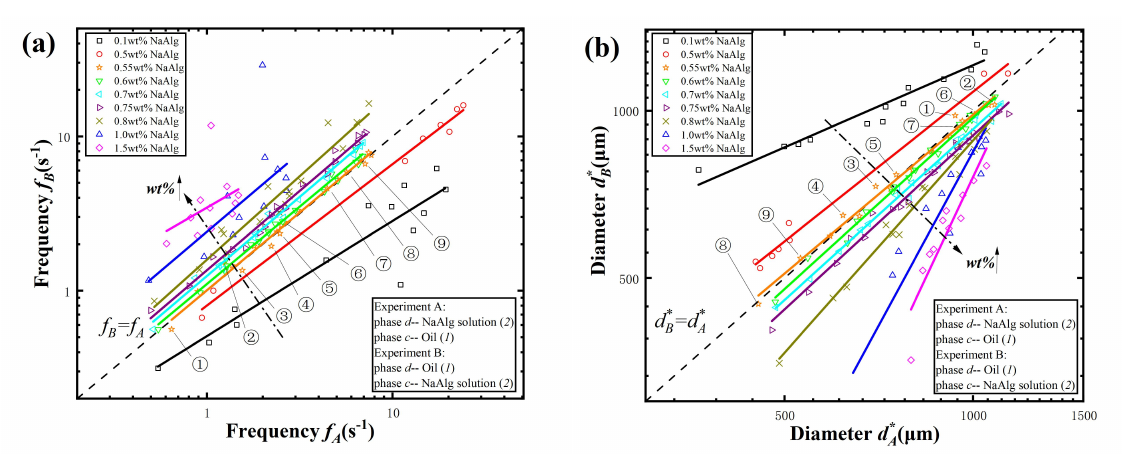}
	  \caption{Comparative graphs detailing the frequency $f$ and equivalent diameter $d^*$ of two-phase droplets in phase interchanging experiments with various mass fractions of NaAlg solution. NaAlg solution mass fractions used were \SI{0.1}{wt \%}($\square$), \SI{0.5}{wt \%}(\textcolor{red}{$\bigcirc$}), \SI{0.55}{wt \%}(\textcolor{orange}{\ding{73}}), \SI{0.6}{wt \%}(\textcolor{green}{$\bigtriangledown$}), \SI{0.7}{wt \%}(\textcolor{cyan}{$\bigtriangleleft$}), \SI{0.75}{wt \%}(\textcolor{violet}{$\triangleright$}), \SI{0.8}{wt \%}(\textcolor{olive}{$\times$}), \SI{1.0}{wt \%}(\textcolor{blue}{$\bigtriangleup$}), and \SI{1.5}{wt \%}(\textcolor{magenta}{$\lozenge$}). The nine marked points, from \ding{172} to \ding{180}, correspond to the experimental data under diverse two-phase flow conditions. Incremental mass fractions of NaAlg solution are indicated by dotted lines, and differently colored lines denote the fitted data lines for each distinct mass fraction. In Experiment A, NaAlg solution serves as the dispersed phase and oil as the continuous phase. In contrast, oil serves as the dispersed phase and NaAlg solution as the continuous phase in Experiment B. (a) Comparative graph of the two-phase droplet frequency $f$ in phase interchanging experiments, with the dashed line representing a state of equality in this frequency ($f_{B} =f_{A}$). (b) Comparative graph of the two-phase droplet equivalent diameter $d^*$ in phase interchanging experiments, with the dashed line illustrating a state of equality for the two-phase droplet equivalent diameter ($d_{B}^{*} =d_{A}^{*}$).}\label{fig6}
\end{figure*}

The droplet frequencies $f$ and equivalent diameters $d^*$ obtained from both Experiments A and B are compared, when the two-phase flow rates are identical $(Q_c=Q_d)$, as shown in Fig. \ref{fig6}.

As shown in Fig. \ref{fig6}, \SI{0.55}{wt\%} NaAlg solution marks the transition mass fraction threshold for synchronous transformations of temporal and spatial characteristics during phase interchanging experiments. Under this condition, $f$ and $d^*$ of both NaAlg and oil droplets remain constant synchronously following the phase interchanging. The existence of this synchronous transition phenomenon is unrelated to the specific flow rate $Q$, but to the mass fraction of the NaAlg solutions. As shown in Table \ref{tab4}, compare the characteristics of NaAlg droplets with those of the oil droplets. Specifically, when the NaAlg solution mass fraction is lower than \SI{0.55}{wt\%}, $f_A$ is higher than $f_B$ and $d_A^*$ is smaller than $d_B^*$. Conversely, as the NaAlg solution mass fraction increases, $f_A$ diminishes and $d_A^*$ enlarges.

\begin{table*}[width=0.9\textwidth,htbp]
\caption{Comparison of the droplet frequency $f$ and the equivalent diameter $d^*$ during phase transition.}\label{tab4}
\begin{tabular*}{\tblwidth}{@{}LLL@{}}
\hline
Mass fraction                & Droplet frequency $f$ & Droplet equivalent diameter $d^*$   \\ \hline
$0.1 \sim$ \SI{0.5}{wt \%}   & $f_{A} >f_{B}$        & $d_{A}^{*} <d_{B}^{*}$              \\
\SI{0.55}{wt \%}             & $f_{A} =f_{B}$        & $d_{A}^{*} =d_{B}^{*}$              \\
$0.6 \sim$ \SI{1.5}{wt \%}   & $f_{A} <f_{B}$        & $d_{A}^{*} >d_{B}^{*}$              \\ \hline
\end{tabular*}
\end{table*}

The Capillary number $Ca$ signifies the ratio between shear stress and surface tension, the Weber number $We$ encapsulates the ratio between inertia and surface tension, and the Reynolds number $Re$ represents the ratio between inertial and viscous forces. The Carreau model, which is effective in characterizing low viscosity, is utilized in its primary definition for the computation of dimensionless numbers. In consideration of variations in velocity $u$, the dynamic viscosity ${\eta}'$ for non-Newtonian fluids \cite{J1984pre} is calculated as

\begin{equation}\label{equ6}
\begin{aligned}
    \eta'=((1+(\frac{3n+1}{4n}\frac{8u_2}{D}\lambda)^2)^{\frac{n-1}{2}}(\eta_0-\eta_\infty)+\eta_\infty)(\frac{3n+1}{4n})
\end{aligned}
\end{equation}

The velocity $u_2=Q_2/\pi(D_2/2)^2$ is derived from the flow rate $Q_2$. The formulas used to calculate the three prevalent dimensionless numbers for both Newtonian fluids and shear-thinning non-Newtonian fluids \cite{Cerdeira2020rev} are shown in Table \ref{tab5}.

\begin{table*}[width=0.9\textwidth,htbp]
\caption{The formulas for dimensionless numbers used in both Newtonian and non-Newtonian fluids.}\label{tab5}
\begin{tabular*}{\tblwidth}{@{}LLL@{}}
\hline
Dimensionless numbers & Newtonian fluid                        & Non-Newtonian fluid                     \\ \hline
$Ca$                  & $Ca_1=\frac{\eta_1 u_1}{\sigma}$       & $Ca_2=\frac{\eta' u_2}{\sigma}$         \\
$We$                  & $We_1=\frac{\rho_1 D_1 u_1^2}{\sigma}$ & $We_2=\frac{\rho_2 D_2 u_2^2}{\sigma}$  \\
$Re$                  & $Re_1=\frac{\rho_1 D_1 u_1}{\eta_1}$   & $Re_2=\frac{\rho_2 D_2 u_2}{\eta'}$     \\ \hline
\end{tabular*}
\end{table*}

\begin{figure*}[htbp]
	\centering
		\includegraphics[scale=0.7]{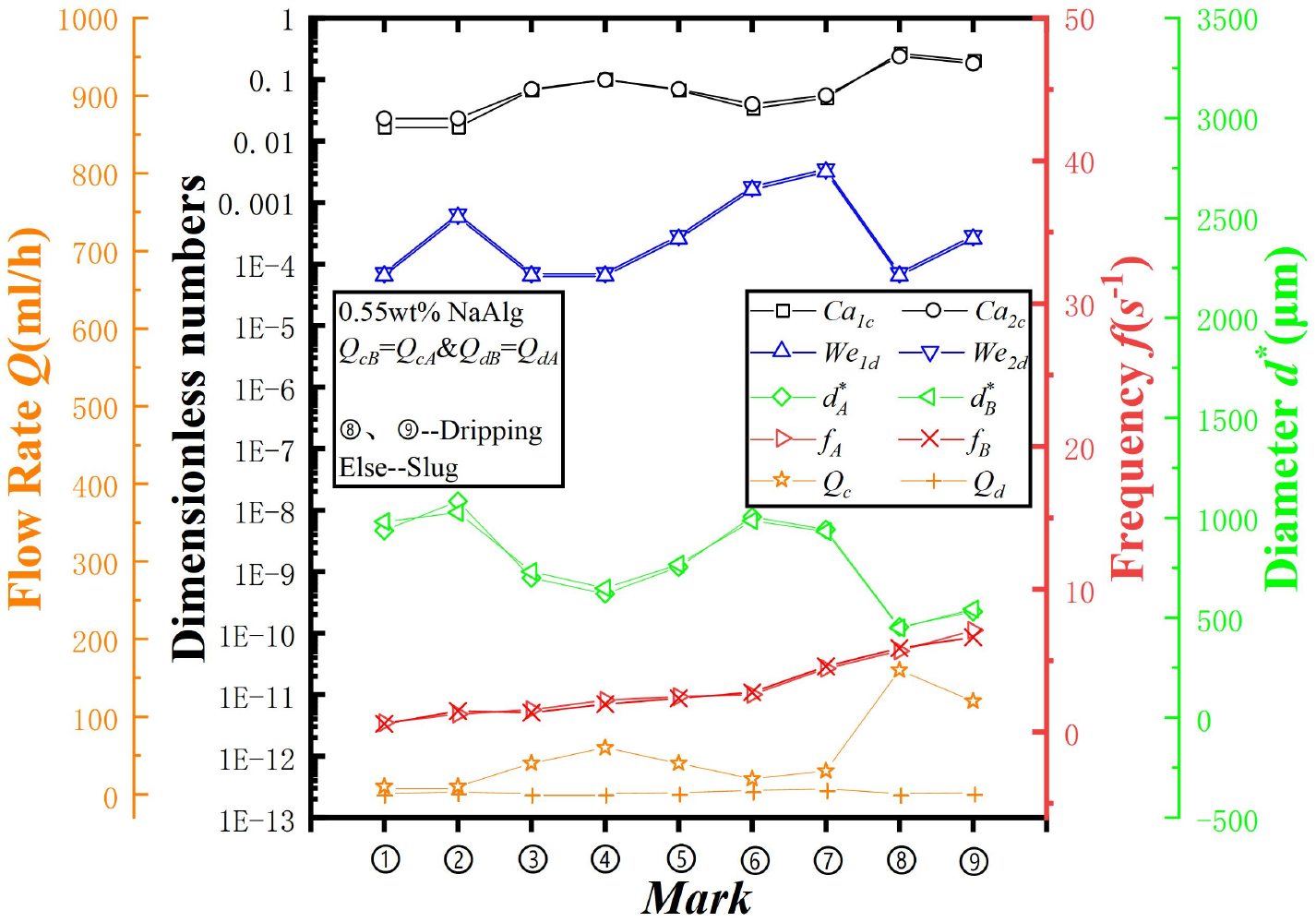}
	  \caption{Comparative graph of experimental parameters in phase interchanging experiments using \SI{0.55}{wt \%} NaAlg solution with identical two phase flow rates ($Q_{cB} =Q_{cA}$, $Q_{dB} =Q_{dA}$). Nine sets of data denoted by markings \ding{172} to \ding{180} is comprised. Notably, marks \ding{179} and \ding{180} generate dripping flow, while the remaining marks result in slug flow. The depicted parameters include the Capillary number of the continuous phase ($Ca_{1c}$ ($\square$), $Ca_{2c}$ ($\bigcirc$)), the Weber number of the dispersed phase ($We_{1d}$ (\textcolor{blue}{$\bigtriangleup$}), $We_{2d}$ (\textcolor{blue}{$\bigtriangledown$})), the droplet equivalent diameter ($d_{A}^{*}$ (\textcolor{green}{$\lozenge$}), $d_{B}^{*}$ (\textcolor{green}{$\bigtriangleleft$})),the droplet frequency ($f_A$ (\textcolor{red}{$\triangleright$}), $f_B$ (\textcolor{red}{$\times$})), and the two phase flow rate ($Q_{c}$ (\textcolor{orange}{\ding{73}}), $Q_{d}$ (\textcolor{orange}{+})).}\label{fig7}
\end{figure*}

As shown in Fig. \ref{fig7}, various parameters derived from the nine two-phase flow conditions (shown in Fig. \ref{fig6}) for the \SI{0.55}{wt\%} NaAlg solution are included.

Fig. \ref{fig7} illustrates that $d^*$ does not correspond with an increase in $f$, instead, its changing pattern mirrors the alterations of the dispersed phase Weber number $We_d$. The two phase flow rate remains consistent $(Q_{cB} =Q_{cA}$, $Q_{dB} =Q_{dA})$, which leads to minimal differences in $f_A$ and $f_B$ and $d_A^*$ and $d_B^*$, thereby maintaining the temporal and spatial characteristics of microdroplets synchronously constant. During instances of low Capillary number $Ca$, surface tension outstrips the shear stress, accounting for the insignificance of surface tension effects. The Weber number $We$ is less than the Capillary number $Ca$, indicating a larger influence of velocity on the dimensionless numbers. When $Q_d$ is held constant, the increase of $Ca_c$ results in a decrease in $d^*$, which suggests that higher $Ca_c$ results in greater shear stress, expediting the fragmentation process of the dispersed phase droplets and leading to the formation of smaller droplets \cite{Liu2018for}. Images and characteristic parameters of the droplets are shown in Table \ref{tab6}, when the \SI{0.55}{wt\%} NaAlg solution functions as the dispersed phase and the continuous phase respectively.

\begin{table*}[width=1\textwidth,htbp]
\caption{Features of \SI{0.55}{wt \%} NaAlg droplets under identical two phase flow rates ($Q_{cB} =Q_{cA}$, $Q_{dB} =Q_{dA}$) in Experiments A and B.}\label{tab6}
\begin{tabular*}{\tblwidth}{@{}CCCCCCCCCCC@{}}
\hline
\multicolumn{2}{c}{Mark}                                                           & \ding{172}  & \ding{173}  & \ding{174}  & \ding{175}  & \ding{176}  & \ding{177}  & \ding{178}  & \ding{179}  & \ding{180}  \\ \hline
\multicolumn{2}{c}{$Q_d(\SI{}{ml/h})$}                                             & 1           & 3           & 1           & 1           & 2           & 5           & 7           & 1           & 2           \\
\multicolumn{2}{c}{$Q_c(\SI{}{ml/h})$}                                             & 10          & 10          & 40          & 60          & 40          & 20          & 30          & 160         & 120         \\ \hline
\multicolumn{2}{c}{Flow pattern}                                                   & Slug        & Slug        & Slug        & Slug        & Slug        & Slug        & Slug        & Dripping    & Dripping    \\ \hline
\multicolumn{2}{c}{\begin{tabular}[c]{@{}c@{}}Image\\ (Experiment A)\end{tabular}} & \begin{minipage}[b]{0.2\columnwidth}
		                                                                              \centering
		                                                                              \raisebox{-.5\height}{\includegraphics[scale=0.4]{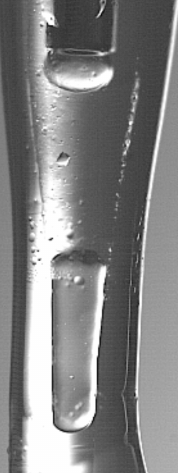}}
	                                                                               \end{minipage}       
                                                                                   & \begin{minipage}[b]{0.2\columnwidth}
		                                                                              \centering
		                                                                              \raisebox{-.5\height}{\includegraphics[scale=0.4]{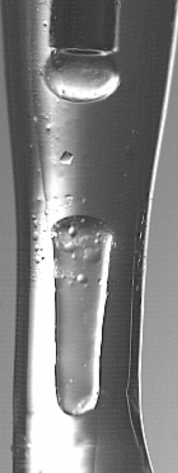}}
	                                                                               \end{minipage}       
                                                                                   & \begin{minipage}[b]{0.2\columnwidth}
		                                                                              \centering
		                                                                              \raisebox{-.5\height}{\includegraphics[scale=0.4]{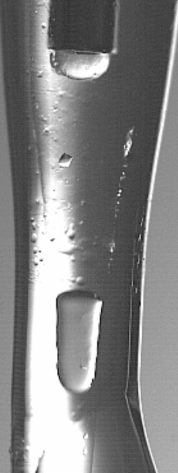}}
	                                                                               \end{minipage}       
                                                                                   & \begin{minipage}[b]{0.2\columnwidth}
		                                                                              \centering
		                                                                              \raisebox{-.5\height}{\includegraphics[scale=0.4]{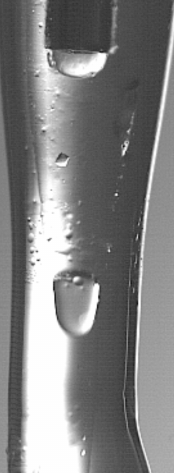}}
	                                                                               \end{minipage}       
                                                                                   & \begin{minipage}[b]{0.2\columnwidth}
		                                                                              \centering
		                                                                              \raisebox{-.5\height}{\includegraphics[scale=0.4]{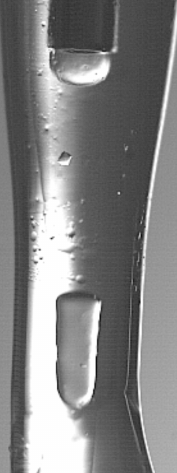}}
	                                                                               \end{minipage}       
                                                                                   & \begin{minipage}[b]{0.2\columnwidth}
		                                                                              \centering
		                                                                              \raisebox{-.5\height}{\includegraphics[scale=0.4]{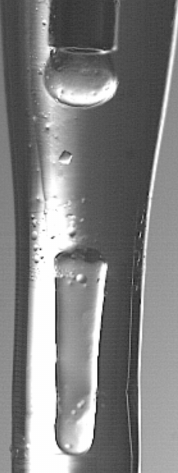}}
	                                                                               \end{minipage}       
                                                                                   & \begin{minipage}[b]{0.2\columnwidth}
		                                                                              \centering
		                                                                              \raisebox{-.5\height}{\includegraphics[scale=0.4]{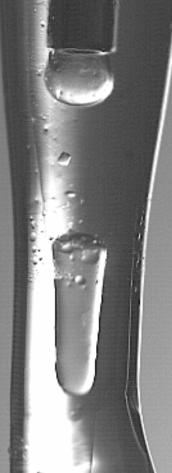}}
	                                                                               \end{minipage}       
                                                                                   & \begin{minipage}[b]{0.2\columnwidth}
		                                                                              \centering
		                                                                              \raisebox{-.5\height}{\includegraphics[scale=0.4]{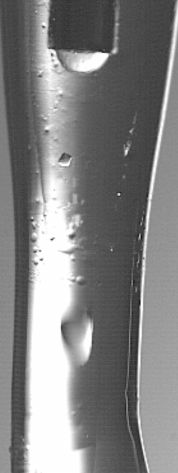}}
	                                                                               \end{minipage}       
                                                                                   & \begin{minipage}[b]{0.2\columnwidth}
		                                                                              \centering
		                                                                              \raisebox{-.5\height}{\includegraphics[scale=0.4]{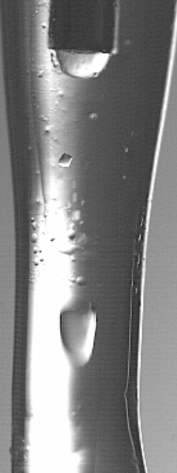}}
	                                                                               \end{minipage}                                                                                                             \\
\multicolumn{2}{c}{\begin{tabular}[c]{@{}c@{}}Image\\ (Experiment B)\end{tabular}} & \begin{minipage}[b]{0.2\columnwidth}
		                                                                              \centering
		                                                                              \raisebox{-.5\height}{\includegraphics[scale=0.4]{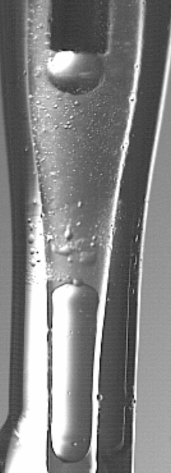}}
	                                                                               \end{minipage}
                                                                                   & \begin{minipage}[b]{0.2\columnwidth}
		                                                                              \centering
		                                                                              \raisebox{-.5\height}{\includegraphics[scale=0.4]{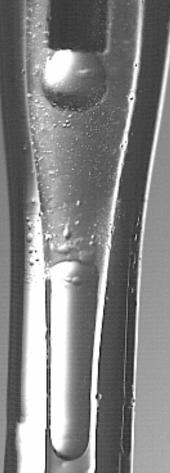}}
	                                                                               \end{minipage}
                                                                                   & \begin{minipage}[b]{0.2\columnwidth}
		                                                                              \centering
		                                                                              \raisebox{-.5\height}{\includegraphics[scale=0.4]{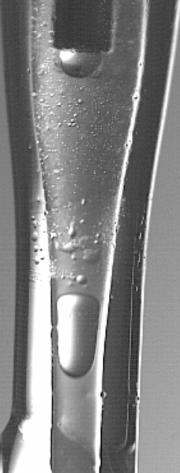}}
	                                                                               \end{minipage}
                                                                                   & \begin{minipage}[b]{0.2\columnwidth}
		                                                                              \centering
		                                                                              \raisebox{-.5\height}{\includegraphics[scale=0.4]{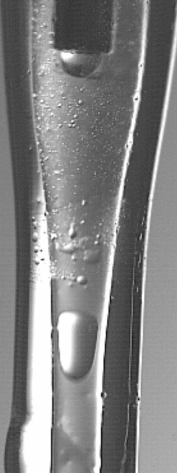}}
	                                                                               \end{minipage}
                                                                                   & \begin{minipage}[b]{0.2\columnwidth}
		                                                                              \centering
		                                                                              \raisebox{-.5\height}{\includegraphics[scale=0.4]{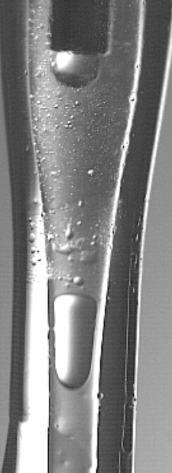}}
	                                                                               \end{minipage}
                                                                                   & \begin{minipage}[b]{0.2\columnwidth}
		                                                                              \centering
		                                                                              \raisebox{-.5\height}{\includegraphics[scale=0.4]{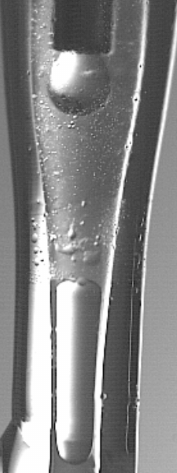}}
	                                                                               \end{minipage}
                                                                                   & \begin{minipage}[b]{0.2\columnwidth}
		                                                                              \centering
		                                                                              \raisebox{-.5\height}{\includegraphics[scale=0.4]{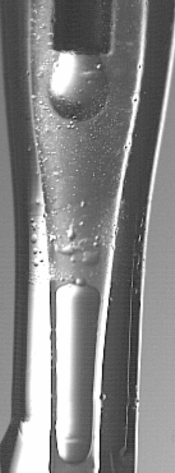}}
	                                                                               \end{minipage}
                                                                                   & \begin{minipage}[b]{0.2\columnwidth}
		                                                                              \centering
		                                                                              \raisebox{-.5\height}{\includegraphics[scale=0.4]{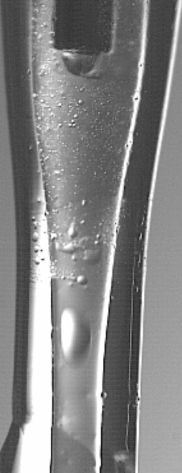}}
	                                                                               \end{minipage}
                                                                                   & \begin{minipage}[b]{0.2\columnwidth}
		                                                                              \centering
		                                                                              \raisebox{-.5\height}{\includegraphics[scale=0.4]{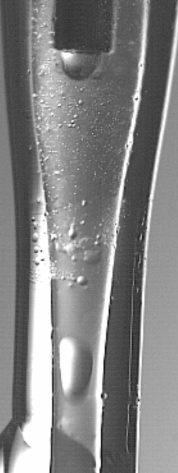}}
	                                                                               \end{minipage}                                                                                                             \\ \hline
\end{tabular*}
\end{table*}

\section{Conclusions}\label{sec4}

In the present paper, the temporal and spatial behaviors of non-Newtonian and Newtonian fluids liquid-liquid system with interchanging two phases in a converging coaxial microchannel has been studied. Two sets of experiments were conducted: Experiment A with NaAlg solution as the dispersed phase and soybean oil as the continuous phase, and Experiment B with soybean oil as the dispersed phase and NaAlg solution as the continuous phase. Three different flow patterns have been observed: slug, dripping, and jetting flow. Phase diagrams have been plotted in flow rate ratio $Q_d / Q_c$, the non-Newtonian index $n$, dimensionless droplet frequency $f \cdot \tau$, and dimensionless equivalent diameter $d^* / D_c$. The transition between the flow regimes has been found to be mainly described by $n$. Based on the results of this study, the following conclusions can be drawn:

(1) When NaAlg solution has high mass fractions (\SI{1.0}{wt\%} and \SI{1.5}{wt\%}), the Carreau model exhibits some deviations. The Carreau model is modified to capture the shear-thinning behavior and the non-Newtonian index $n$ is obtained. It is found that for higher viscosity, the non-Newtonian index can be $n<0$, which is rarely reported, contributing to phase diagram separation and flow pattern classification.

(2) $f \cdot \tau \sim Q_d / Q_c$ and $d^* / D_c \sim Q_d / Q_c$ phase diagrams are plotted to verify the importance of $n$. Without the introduction of $n$, boundaries between flow patterns are not clear, and temporal and spatial behaviors are difficult to analyze.

(3) $f \cdot \tau \sim\left(Q_d / Q_c\right)^n$ and $d^* / D_c \sim\left(Q_d / Q_c\right)^n$ phase diagrams are conducted by introducing $n$. In both Experiment A and Experiment B, clear boundaries between flow patterns are observed. The temporal behavior exhibits as a "butterfly distribution" and the spatial behavior as a "grape distribution". The distribution of flow patterns are divided by $n$, symmetric about $\left(Q_d / Q_c\right)^n=1$, for either $Q_d / Q_c=1$ or $n=0$. As for $-1<n<0$, an increase in $(Q_d / Q_c)^n$ leads to an increase in $f \cdot \tau$ and a decrease in $d^* / D_c$; and as for $0<n<1$, the pattern is opposite.

(4) A synchronous transition phenomenon (\SI{0.55}{wt\%} NaAlg solution) is identified when the two-phase flow rates remain constant $(Q_{cB}=Q_{cA},Q_{dB}=Q_{dA})$ during phase interchanging. At this phenomenon shows, the temporal and spatial characteristics of microdroplets remain unchanged $(f_A=f_B,d_A^*=d_B^*)$ after interchanging of disperse and continuous phases. As for lower mass fractions of NaAlg solution (\SI{0.1}{wt\%}, \SI{0.5}{wt\%}), $d_A^*\leq d_B^*$ and $f_{A}\geq f_{B}$; and as for higher mass fractions (\SI{1.0}{wt\%}, \SI{1.5}{wt\%}), $d_{A}^{*}\geq d_{B}^{*}$ and $f_{A}\leq f_{B}$.

The modification of the Carreau model enables the non-Newtonian index $n$ to have physical significance when it is negative. By conducting phase interchanging experiments in a converging coaxial microchannel, these findings are expected to provide better understanding on the temporal and spatial behaviors of non-Newtonian monodisperse microdroplets.

\section*{Declaration of Competing Interest}

The authors declare that they have no known competing financial interests or personal relationships that could have appeared to influence the work reported in this paper.

\section*{Acknowledgements}

The author acknowledges the Natural Science Foundation of China (No. 11832017 and 11772183).

\printcredits

\bibliographystyle{model1-num-names}

\bibliography{ref}

\begin{thebibliography}{56}
\expandafter\ifx\csname natexlab\endcsname\relax\def\natexlab#1{#1}\fi
\providecommand{\url}[1]{\texttt{#1}}
\providecommand{\href}[2]{#2}
\providecommand{\path}[1]{#1}
\providecommand{\DOIprefix}{doi:}
\providecommand{\ArXivprefix}{arXiv:}
\providecommand{\URLprefix}{URL: }
\providecommand{\Pubmedprefix}{pmid:}
\providecommand{\doi}[1]{\href{http://dx.doi.org/#1}{\path{#1}}}
\providecommand{\Pubmed}[1]{\href{pmid:#1}{\path{#1}}}
\providecommand{\bibinfo}[2]{#2}
\ifx\xfnm\relax \def\xfnm[#1]{\unskip,\space#1}\fi
\bibitem[{Lari et~al.(2021)Lari, Zahedi, Ghourchian, and Khatibi}]{Lari2021mic}
\bibinfo{author}{A.~S. Lari}, \bibinfo{author}{P.~Zahedi},
  \bibinfo{author}{H.~Ghourchian}, \bibinfo{author}{A.~Khatibi},
\newblock \bibinfo{title}{Microfluidic-based synthesized carboxymethyl chitosan
  nanoparticles containing metformin for diabetes therapy: In vitro and in vivo
  assessments},
\newblock \bibinfo{journal}{Carbohydrate Polymers} \bibinfo{volume}{261}
  (\bibinfo{year}{2021}).
\bibitem[{Su et~al.(2021)Su, Liang, and Tan}]{Su2021mic}
\bibinfo{author}{W.~Su}, \bibinfo{author}{D.~Liang}, \bibinfo{author}{M.~Tan},
\newblock \bibinfo{title}{Microfluidic strategies for sample separation and
  rapid detection of food allergens},
\newblock \bibinfo{journal}{Trends in Food Science \& Technology}
  \bibinfo{volume}{110} (\bibinfo{year}{2021}) \bibinfo{pages}{213--225}.
\bibitem[{Tan et~al.(2021)Tan, Powles, Zhang, and Shen}]{Tan2021go}
\bibinfo{author}{W.~Tan}, \bibinfo{author}{E.~Powles},
  \bibinfo{author}{L.~Zhang}, \bibinfo{author}{W.~Shen},
\newblock \bibinfo{title}{Go with the capillary flow. simple thread-based
  microfluidics},
\newblock \bibinfo{journal}{Sensors and Actuators B-Chemical}
  \bibinfo{volume}{334} (\bibinfo{year}{2021}).
\bibitem[{Huang et~al.(2018)Huang, Chou, Chen, Chou, Wu, Chen, and
  Lee}]{Huang2018an}
\bibinfo{author}{W.-Y. Huang}, \bibinfo{author}{S.-T. Chou},
  \bibinfo{author}{C.-H. Chen}, \bibinfo{author}{S.-Y. Chou},
  \bibinfo{author}{J.-H. Wu}, \bibinfo{author}{Y.-C. Chen},
  \bibinfo{author}{G.-B. Lee},
\newblock \bibinfo{title}{An automatic integrated microfluidic system for
  allergy microarray chips},
\newblock \bibinfo{journal}{Analyst} \bibinfo{volume}{143}
  (\bibinfo{year}{2018}) \bibinfo{pages}{2285--2292}.
\bibitem[{Zheng et~al.(2003)Zheng, Roach, and Ismagilov}]{Zheng2003scr}
\bibinfo{author}{B.~Zheng}, \bibinfo{author}{L.~S. Roach},
  \bibinfo{author}{R.~F. Ismagilov},
\newblock \bibinfo{title}{Screening of protein crystallization conditions on a
  microfluidic chip using nanoliter-size droplets},
\newblock \bibinfo{journal}{Journal of the American Chemical Society}
  \bibinfo{volume}{125} (\bibinfo{year}{2003}) \bibinfo{pages}{11170--11171}.
\bibitem[{Hung et~al.(2006)Hung, Choi, Tseng, Tan, Shea, and Lee}]{Hung2006alt}
\bibinfo{author}{L.~H. Hung}, \bibinfo{author}{K.~M. Choi},
  \bibinfo{author}{W.~Y. Tseng}, \bibinfo{author}{Y.~C. Tan},
  \bibinfo{author}{K.~J. Shea}, \bibinfo{author}{A.~P. Lee},
\newblock \bibinfo{title}{Alternating droplet generation and controlled dynamic
  droplet fusion in microfluidic device for cds nanoparticle synthesis},
\newblock \bibinfo{journal}{Lab on a Chip} \bibinfo{volume}{6}
  (\bibinfo{year}{2006}) \bibinfo{pages}{174--178}.
\bibitem[{Zhai et~al.(2020)Zhai, Meng, Yang, Zhang, and Jin}]{Zhai2020det}
\bibinfo{author}{L.~Zhai}, \bibinfo{author}{Z.~Meng},
  \bibinfo{author}{J.~Yang}, \bibinfo{author}{H.~Zhang},
  \bibinfo{author}{N.~Jin},
\newblock \bibinfo{title}{Detection of interfacial structures in inclined
  liquid-liquid flows using parallel-wire array probe and planar laser-induced
  fluorescence methods},
\newblock \bibinfo{journal}{Sensors} \bibinfo{volume}{20}
  (\bibinfo{year}{2020}).
\bibitem[{Qian et~al.(2019)Qian, Li, Wu, Jin, and Sunden}]{Qian2019a}
\bibinfo{author}{J.-y. Qian}, \bibinfo{author}{X.-j. Li},
  \bibinfo{author}{Z.~Wu}, \bibinfo{author}{Z.-j. Jin},
  \bibinfo{author}{B.~Sunden},
\newblock \bibinfo{title}{A comprehensive review on liquid-liquid two-phase
  flow in microchannel: flow pattern and mass transfer},
\newblock \bibinfo{journal}{Microfluidics and Nanofluidics}
  \bibinfo{volume}{23} (\bibinfo{year}{2019}).
\bibitem[{Xu et~al.(2009)Xu, Hashimoto, Dang, Hoare, Kohane, Whitesides,
  Langer, and Anderson}]{Xu2009pre}
\bibinfo{author}{Q.~Xu}, \bibinfo{author}{M.~Hashimoto}, \bibinfo{author}{T.~T.
  Dang}, \bibinfo{author}{T.~Hoare}, \bibinfo{author}{D.~S. Kohane},
  \bibinfo{author}{G.~M. Whitesides}, \bibinfo{author}{R.~Langer},
  \bibinfo{author}{D.~G. Anderson},
\newblock \bibinfo{title}{Preparation of monodisperse biodegradable polymer
  microparticles using a microfluidic flow-focusing device for controlled drug
  delivery},
\newblock \bibinfo{journal}{Small} \bibinfo{volume}{5} (\bibinfo{year}{2009})
  \bibinfo{pages}{1575--1581}.
\bibitem[{Zhao et~al.(2009)Zhao, Pan, Zhang, Guo, Liu, Wang, Chen, Zhao, and
  Chan}]{Zhao2009gen}
\bibinfo{author}{L.~B. Zhao}, \bibinfo{author}{L.~Pan},
  \bibinfo{author}{K.~Zhang}, \bibinfo{author}{S.~S. Guo},
  \bibinfo{author}{W.~Liu}, \bibinfo{author}{Y.~Wang},
  \bibinfo{author}{Y.~Chen}, \bibinfo{author}{X.~Z. Zhao},
  \bibinfo{author}{H.~L.~W. Chan},
\newblock \bibinfo{title}{Generation of janus alginate hydrogel particles with
  magnetic anisotropy for cell encapsulation},
\newblock \bibinfo{journal}{Lab on a Chip} \bibinfo{volume}{9}
  (\bibinfo{year}{2009}) \bibinfo{pages}{2981--2986}.
\bibitem[{Zheng et~al.(2004)Zheng, Tice, and Ismagilov}]{Zheng2004for}
\bibinfo{author}{B.~Zheng}, \bibinfo{author}{J.~D. Tice},
  \bibinfo{author}{R.~F. Ismagilov},
\newblock \bibinfo{title}{Formation of arrayed droplets of soft lithography and
  two-phase fluid flow, and application in protein crystallization},
\newblock \bibinfo{journal}{Advanced Materials} \bibinfo{volume}{16}
  (\bibinfo{year}{2004}) \bibinfo{pages}{1365--1368}.
\bibitem[{Fu et~al.(2016)Fu, Ma, and Li}]{Fu2016bre}
\bibinfo{author}{T.~Fu}, \bibinfo{author}{Y.~Ma}, \bibinfo{author}{H.~Z. Li},
\newblock \bibinfo{title}{Breakup dynamics of slender droplet formation in
  shear-thinning fluids in flow-focusing devices},
\newblock \bibinfo{journal}{Chemical Engineering Science} \bibinfo{volume}{144}
  (\bibinfo{year}{2016}) \bibinfo{pages}{75--86}.
\bibitem[{Kovalev et~al.(2018)Kovalev, Yagodnitsyna, and
  Bilsky}]{Kovalev2018flow}
\bibinfo{author}{A.~V. Kovalev}, \bibinfo{author}{A.~A. Yagodnitsyna},
  \bibinfo{author}{A.~V. Bilsky},
\newblock \bibinfo{title}{Flow hydrodynamics of immiscible liquids with low
  viscosity ratio in a rectangular microchannel with t-junction},
\newblock \bibinfo{journal}{Chemical Engineering Journal} \bibinfo{volume}{352}
  (\bibinfo{year}{2018}) \bibinfo{pages}{120--132}.
\bibitem[{Timung et~al.(2015)Timung, Tiwari, Singh, Mandal, and
  Bandyopadhyay}]{Timung2015cap}
\bibinfo{author}{S.~Timung}, \bibinfo{author}{V.~Tiwari},
  \bibinfo{author}{A.~K. Singh}, \bibinfo{author}{T.~K. Mandal},
  \bibinfo{author}{D.~Bandyopadhyay},
\newblock \bibinfo{title}{Capillary force mediated flow patterns and
  non-monotonic pressure drop characteristics of oil-water microflows},
\newblock \bibinfo{journal}{Canadian Journal of Chemical Engineering}
  \bibinfo{volume}{93} (\bibinfo{year}{2015}) \bibinfo{pages}{1736--1743}.
\bibitem[{Dang et~al.(2013)Dang, Yue, Chen, and Yuan}]{Dang2013for}
\bibinfo{author}{M.~Dang}, \bibinfo{author}{J.~Yue}, \bibinfo{author}{G.~Chen},
  \bibinfo{author}{Q.~Yuan},
\newblock \bibinfo{title}{Formation characteristics of taylor bubbles in a
  microchannel with a converging shape mixing junction},
\newblock \bibinfo{journal}{Chemical Engineering Journal} \bibinfo{volume}{223}
  (\bibinfo{year}{2013}) \bibinfo{pages}{99--109}.
\bibitem[{Yin et~al.(2018)Yin, Pei, Peng, Zhang, and
  Srinivasakannan}]{Yin2018stu}
\bibinfo{author}{S.~Yin}, \bibinfo{author}{J.~Pei}, \bibinfo{author}{J.~Peng},
  \bibinfo{author}{L.~Zhang}, \bibinfo{author}{C.~Srinivasakannan},
\newblock \bibinfo{title}{Study on mass transfer behavior of extracting la(iii)
  with ehehpa (p507) using rectangular cross-section microchannel},
\newblock \bibinfo{journal}{Hydrometallurgy} \bibinfo{volume}{175}
  (\bibinfo{year}{2018}) \bibinfo{pages}{64--69}.
\bibitem[{Sontti and Atta(2018)}]{Sontti2018for}
\bibinfo{author}{S.~G. Sontti}, \bibinfo{author}{A.~Atta},
\newblock \bibinfo{title}{Formation characteristics of taylor bubbles in
  power-law liquids flowing through a microfluidic co-flow device},
\newblock \bibinfo{journal}{Journal of Industrial and Engineering Chemistry}
  \bibinfo{volume}{65} (\bibinfo{year}{2018}) \bibinfo{pages}{82--94}.
\bibitem[{Deng et~al.(2017)Deng, Wang, Huang, and Cheng}]{Deng2017num}
\bibinfo{author}{C.~Deng}, \bibinfo{author}{H.~Wang},
  \bibinfo{author}{W.~Huang}, \bibinfo{author}{S.~Cheng},
\newblock \bibinfo{title}{Numerical and experimental study of oil-in-water
  (o/w) droplet formation in a co-flowing capillary device},
\newblock \bibinfo{journal}{Colloids and Surfaces a-Physicochemical and
  Engineering Aspects} \bibinfo{volume}{533} (\bibinfo{year}{2017})
  \bibinfo{pages}{1--8}.
\bibitem[{Wang(2015)}]{Wang2015spe}
\bibinfo{author}{Z.~L. Wang},
\newblock \bibinfo{title}{Speed up bubbling in a tapered co-flow geometry},
\newblock \bibinfo{journal}{Chemical Engineering Journal} \bibinfo{volume}{263}
  (\bibinfo{year}{2015}) \bibinfo{pages}{346--355}.
\bibitem[{Wang(2022)}]{Wang2022uni}
\bibinfo{author}{Z.~L. Wang},
\newblock \bibinfo{title}{Universal self-scalings in a micro-co-flowing},
\newblock \bibinfo{journal}{Chemical Engineering Science} \bibinfo{volume}{262}
  (\bibinfo{year}{2022}).
\bibitem[{Wu et~al.(2015)Wu, Luo, Liu, Li, Chen, Feng, and He}]{Wu2015dra}
\bibinfo{author}{P.~Wu}, \bibinfo{author}{Z.~Luo}, \bibinfo{author}{Z.~Liu},
  \bibinfo{author}{Z.~Li}, \bibinfo{author}{C.~Chen},
  \bibinfo{author}{L.~Feng}, \bibinfo{author}{L.~He},
\newblock \bibinfo{title}{Drag-induced breakup mechanism for droplet generation
  in dripping within flow focusing microfluidics},
\newblock \bibinfo{journal}{Chinese Journal of Chemical Engineering}
  \bibinfo{volume}{23} (\bibinfo{year}{2015}) \bibinfo{pages}{7--14}.
\bibitem[{Chen and Ren(2017)}]{Chen2017exp}
\bibinfo{author}{X.~Chen}, \bibinfo{author}{C.~L. Ren},
\newblock \bibinfo{title}{Experimental study on droplet generation in flow
  focusing devices considering a stratified flow with viscosity contrast},
\newblock \bibinfo{journal}{Chemical Engineering Science} \bibinfo{volume}{163}
  (\bibinfo{year}{2017}) \bibinfo{pages}{1--10}.
\bibitem[{Wang et~al.(2021)Wang, Su, Feng, Wang, and Song}]{Wang2021exp}
\bibinfo{author}{Q.~Wang}, \bibinfo{author}{X.~Su}, \bibinfo{author}{Y.~Feng},
  \bibinfo{author}{H.~Wang}, \bibinfo{author}{J.~Song},
\newblock \bibinfo{title}{Experimental study of gas-water digital microflow
  patterns in fracture: Implication for enhancing coalbed methane production},
\newblock \bibinfo{journal}{Journal of Petroleum Science and Engineering}
  \bibinfo{volume}{207} (\bibinfo{year}{2021}).
\bibitem[{Lee and Lee(2021)}]{Lee2021sur}
\bibinfo{author}{S.~Lee}, \bibinfo{author}{J.~Lee},
\newblock \bibinfo{title}{Surface hydrophobicity change on adiabatic two-phase
  flow pattern transitions in horizontal tubes},
\newblock \bibinfo{journal}{Experimental Thermal and Fluid Science}
  \bibinfo{volume}{120} (\bibinfo{year}{2021}).
\bibitem[{Zhang et~al.(2020)Zhang, Wang, Liu, Yi, and Wang}]{Zhang2020exp}
\bibinfo{author}{J.~Zhang}, \bibinfo{author}{C.~Wang},
  \bibinfo{author}{X.~Liu}, \bibinfo{author}{C.~Yi}, \bibinfo{author}{Z.~L.
  Wang},
\newblock \bibinfo{title}{Experimental studies of microchannel tapering on
  droplet forming acceleration in liquid paraffin/ethanol coaxial flows},
\newblock \bibinfo{journal}{Materials} \bibinfo{volume}{13}
  (\bibinfo{year}{2020}).
\bibitem[{Verma and Ghosh(2020)}]{Verma2020eff}
\bibinfo{author}{R.~K. Verma}, \bibinfo{author}{S.~Ghosh},
\newblock \bibinfo{title}{Effect of phase properties on liquid-liquid two-phase
  flow patterns and pressure drop in serpentine mini geometry},
\newblock \bibinfo{journal}{Chemical Engineering Journal} \bibinfo{volume}{397}
  (\bibinfo{year}{2020}).
\bibitem[{Cerdeira et~al.(2020)Cerdeira, Campos, Miranda, and
  Araujo}]{Cerdeira2020rev}
\bibinfo{author}{A.~T.~S. Cerdeira}, \bibinfo{author}{J.~B. L.~M. Campos},
  \bibinfo{author}{J.~M. Miranda}, \bibinfo{author}{J.~D.~P. Araujo},
\newblock \bibinfo{title}{Review on microbubbles and microdroplets flowing
  through microfluidic geometrical elements},
\newblock \bibinfo{journal}{Micromachines} \bibinfo{volume}{11}
  (\bibinfo{year}{2020}).
\bibitem[{Lakzian and Akbarzadeh(2020)}]{Lakzian2020num}
\bibinfo{author}{E.~Lakzian}, \bibinfo{author}{P.~Akbarzadeh},
\newblock \bibinfo{title}{Numerical investigation of unsteady pulsatile
  newtonian/non-newtonian blood flow through curved stenosed arteries},
\newblock \bibinfo{journal}{Bio-Medical Materials and Engineering}
  \bibinfo{volume}{30} (\bibinfo{year}{2020}) \bibinfo{pages}{525--540}.
\bibitem[{Picchi et~al.(2018)Picchi, Barmak, Ullmann, and
  Brauner}]{Picchi2018sta}
\bibinfo{author}{D.~Picchi}, \bibinfo{author}{I.~Barmak},
  \bibinfo{author}{A.~Ullmann}, \bibinfo{author}{N.~Brauner},
\newblock \bibinfo{title}{Stability of stratified two-phase channel flows of
  newtonian/non-newtonian shear-thinning fluids},
\newblock \bibinfo{journal}{International Journal of Multiphase Flow}
  \bibinfo{volume}{99} (\bibinfo{year}{2018}) \bibinfo{pages}{111--131}.
\bibitem[{Xie et~al.(2012)Xie, Halley, and Averous}]{Xie2012rhe}
\bibinfo{author}{F.~Xie}, \bibinfo{author}{P.~J. Halley},
  \bibinfo{author}{L.~Averous},
\newblock \bibinfo{title}{Rheology to understand and optimize processibility,
  structures and properties of starch polymeric materials},
\newblock \bibinfo{journal}{Progress in Polymer Science} \bibinfo{volume}{37}
  (\bibinfo{year}{2012}) \bibinfo{pages}{595--623}.
\bibitem[{Schneider and Gerber(2020)}]{Schneider2020rhe}
\bibinfo{author}{N.~Schneider}, \bibinfo{author}{M.~Gerber},
\newblock \bibinfo{title}{Rheological properties of digestate from agricultural
  biogas plants: An overview of measurement techniques and influencing
  factors},
\newblock \bibinfo{journal}{Renewable \& Sustainable Energy Reviews}
  \bibinfo{volume}{121} (\bibinfo{year}{2020}).
\bibitem[{Zhao and Middelberg(2011)}]{Zhao2011two}
\bibinfo{author}{C.-X. Zhao}, \bibinfo{author}{A.~P.~J. Middelberg},
\newblock \bibinfo{title}{Two-phase microfluidic flows},
\newblock \bibinfo{journal}{Chemical Engineering Science} \bibinfo{volume}{66}
  (\bibinfo{year}{2011}) \bibinfo{pages}{1394--1411}.
\bibitem[{Granados-Ortiz et~al.(2021)Granados-Ortiz, Jimenez-Salas, and
  Ortega-Casanova}]{Granados2021app}
\bibinfo{author}{F.~J. Granados-Ortiz}, \bibinfo{author}{M.~Jimenez-Salas},
  \bibinfo{author}{J.~Ortega-Casanova},
\newblock \bibinfo{title}{Application of shear-thinning and shear-thickening
  fluids to computational fluid mechanics of high-reynolds impinging turbulent
  jets for cooling engineering},
\newblock \bibinfo{journal}{International Journal of Thermal Sciences}
  \bibinfo{volume}{162} (\bibinfo{year}{2021}).
\bibitem[{Huang et~al.(2019)Huang, Su, and Lee}]{Huang2019on}
\bibinfo{author}{Y.~N. Huang}, \bibinfo{author}{W.~D. Su},
  \bibinfo{author}{C.~B. Lee},
\newblock \bibinfo{title}{On the weissenberg effect of turbulence},
\newblock \bibinfo{journal}{Theoretical and Applied Mechanics Letters}
  \bibinfo{volume}{9} (\bibinfo{year}{2019}) \bibinfo{pages}{236--245}.
\bibitem[{Sokhal et~al.(2018)Sokhal, Gangacharyulu, and
  Bulasara}]{Sokhal2018eff}
\bibinfo{author}{K.~S. Sokhal}, \bibinfo{author}{D.~Gangacharyulu},
  \bibinfo{author}{V.~K. Bulasara},
\newblock \bibinfo{title}{Effect of guar gum and salt concentrations on drag
  reduction and shear degradation properties of turbulent flow of water in a
  pipe},
\newblock \bibinfo{journal}{Carbohydrate Polymers} \bibinfo{volume}{181}
  (\bibinfo{year}{2018}) \bibinfo{pages}{1017--1025}.
\bibitem[{Fu et~al.(2015)Fu, Wei, Zhu, and Ma}]{Fu2015flow}
\bibinfo{author}{T.~Fu}, \bibinfo{author}{L.~Wei}, \bibinfo{author}{C.~Zhu},
  \bibinfo{author}{Y.~Ma},
\newblock \bibinfo{title}{Flow patterns of liquid-liquid two-phase flow in
  non-newtonian fluids in rectangular microchannels},
\newblock \bibinfo{journal}{Chemical Engineering and Processing-Process
  Intensification} \bibinfo{volume}{91} (\bibinfo{year}{2015})
  \bibinfo{pages}{114--120}.
\bibitem[{Agarwal et~al.(2020)Agarwal, Singh, Bahga, and
  Gupta}]{Agarwal2020dyn}
\bibinfo{author}{V.~G. Agarwal}, \bibinfo{author}{R.~Singh},
  \bibinfo{author}{S.~S. Bahga}, \bibinfo{author}{A.~Gupta},
\newblock \bibinfo{title}{Dynamics of droplet formation and flow regime
  transition in a t-shaped microfluidic device with a shear-thinning continuous
  phase},
\newblock \bibinfo{journal}{Physical Review Fluids} \bibinfo{volume}{5}
  (\bibinfo{year}{2020}).
\bibitem[{Dziubinski et~al.(2004)Dziubinski, Fidos, and
  Sosno}]{Dziubinski2004the}
\bibinfo{author}{M.~Dziubinski}, \bibinfo{author}{H.~Fidos},
  \bibinfo{author}{M.~Sosno},
\newblock \bibinfo{title}{The flow pattern map of a two-phase non-newtonian
  liquid-gas flow in the vertical pipe},
\newblock \bibinfo{journal}{International Journal of Multiphase Flow}
  \bibinfo{volume}{30} (\bibinfo{year}{2004}) \bibinfo{pages}{551--563}.
\bibitem[{Vagner et~al.(2018)Vagner, Patlazhan, Serra, and Iop}]{Vagner2017for}
\bibinfo{author}{S.~A. Vagner}, \bibinfo{author}{S.~A. Patlazhan},
  \bibinfo{author}{C.~A. Serra}, \bibinfo{author}{Iop},
\newblock \bibinfo{title}{Formation of microdroplets in newtonian and shear
  thinning fluids flowing in coaxial capillaries: Numerical modeling},
\newblock in: \bibinfo{booktitle}{32nd International Conference on Interaction
  of Intense Energy Fluxes with Matter (ELBRUS)}, volume \bibinfo{volume}{946}
  of \textit{\bibinfo{series}{Journal of Physics Conference Series}},
  \bibinfo{year}{2018}, pp. \bibinfo{pages}{1--6}. \URLprefix \url{<Go to
  ISI>://WOS:000446782200117}. \DOIprefix\doi{10.1088/1742-6596/946/1/012117}.
\bibitem[{Taassob et~al.(2017)Taassob, Manshadi, Bordbar, and
  Kamali}]{Taassob2017mon}
\bibinfo{author}{A.~Taassob}, \bibinfo{author}{M.~K.~D. Manshadi},
  \bibinfo{author}{A.~Bordbar}, \bibinfo{author}{R.~Kamali},
\newblock \bibinfo{title}{Monodisperse non-newtonian micro-droplet generation
  in a co-flow device},
\newblock \bibinfo{journal}{Journal of the Brazilian Society of Mechanical
  Sciences and Engineering} \bibinfo{volume}{39} (\bibinfo{year}{2017})
  \bibinfo{pages}{2013--2021}.
\bibitem[{Khater et~al.(2020)Khater, Abdelrehim, Mohammadi, Azarmanesh,
  Jonmaleki, Salahandish, Mohamad, and Sanati-Nezhad}]{Khater2020pic}
\bibinfo{author}{A.~Khater}, \bibinfo{author}{O.~Abdelrehim},
  \bibinfo{author}{M.~Mohammadi}, \bibinfo{author}{M.~Azarmanesh},
  \bibinfo{author}{M.~Jonmaleki}, \bibinfo{author}{R.~Salahandish},
  \bibinfo{author}{A.~Mohamad}, \bibinfo{author}{A.~Sanati-Nezhad},
\newblock \bibinfo{title}{Picoliter agar droplet breakup in microfluidics meets
  microbiology application: numerical and experimental approaches},
\newblock \bibinfo{journal}{Lab on a Chip} \bibinfo{volume}{20}
  (\bibinfo{year}{2020}) \bibinfo{pages}{2175--2187}.
\bibitem[{Bai et~al.(2021)Bai, Zhang, Li, Li, and Joo}]{Bai2021gen}
\bibinfo{author}{F.~Bai}, \bibinfo{author}{H.~Zhang}, \bibinfo{author}{X.~Li},
  \bibinfo{author}{F.~Li}, \bibinfo{author}{S.~W. Joo},
\newblock \bibinfo{title}{Generation and dynamics of janus droplets in
  shear-thinning fluid flow in a double y-type microchannel},
\newblock \bibinfo{journal}{Micromachines} \bibinfo{volume}{12}
  (\bibinfo{year}{2021}).
\bibitem[{Battat et~al.(2022)Battat, Weitz, and Whitesides}]{Battat2022non}
\bibinfo{author}{S.~Battat}, \bibinfo{author}{D.~A. Weitz},
  \bibinfo{author}{G.~M. Whitesides},
\newblock \bibinfo{title}{Nonlinear phenomena in microfluidics},
\newblock \bibinfo{journal}{Chemical Reviews} \bibinfo{volume}{122}
  (\bibinfo{year}{2022}) \bibinfo{pages}{6921--6937}.
\bibitem[{Kumar et~al.(2020)Kumar, Kumar, and Singh}]{Kumar2020inf}
\bibinfo{author}{S.~Kumar}, \bibinfo{author}{V.~Kumar}, \bibinfo{author}{A.~K.
  Singh},
\newblock \bibinfo{title}{Influence of lubricants on the performance of journal
  bearings - a review},
\newblock \bibinfo{journal}{Tribology-Materials Surfaces \& Interfaces}
  \bibinfo{volume}{14} (\bibinfo{year}{2020}) \bibinfo{pages}{67--78}.
\bibitem[{Kokini et~al.(1984)Kokini, Bistany, and L.}]{Kokini1984pre}
\bibinfo{author}{J.~L. Kokini}, \bibinfo{author}{K.~L. Bistany},
  \bibinfo{author}{M.~P. L.},
\newblock \bibinfo{title}{Predicting steady shear and dynamic viscoelastic
  properties of guar and carrageenan using the bird-carreau constitutive
  model},
\newblock \bibinfo{journal}{Journal of Food Science} \bibinfo{volume}{49}
  (\bibinfo{year}{1984}) \bibinfo{pages}{1569--1572}.
\bibitem[{Waele(1923)}]{Waele1923vis}
\bibinfo{author}{A.~Waele},
\newblock \bibinfo{title}{Viscometry and plastometry},
\newblock \bibinfo{journal}{Oil Colour Chemists’ Assoc} \bibinfo{volume}{6}
  (\bibinfo{year}{1923}) \bibinfo{pages}{33--88}.
\bibitem[{Cross(1965)}]{Cross1965rhe}
\bibinfo{author}{M.~M. Cross},
\newblock \bibinfo{title}{Rheology of non-newtonian fluids: a new flow equation
  for pseudoplastic systems},
\newblock \bibinfo{journal}{Journal of Colloid Science} \bibinfo{volume}{20}
  (\bibinfo{year}{1965}) \bibinfo{pages}{417--437}.
\bibitem[{Herschel and Bulkley(1926)}]{Herschel1926kon}
\bibinfo{author}{W.~H. Herschel}, \bibinfo{author}{R.~Bulkley},
\newblock \bibinfo{title}{Konsistenzmessungen von gummi-benzollösungen},
\newblock \bibinfo{journal}{Kolloid-Zeitschrift} \bibinfo{volume}{39}
  (\bibinfo{year}{1926}) \bibinfo{pages}{291--300}.
\bibitem[{Dhiman et~al.(2019)Dhiman, Ghosh, and Baranyi}]{Dhiman2019hyd}
\bibinfo{author}{A.~Dhiman}, \bibinfo{author}{R.~Ghosh},
  \bibinfo{author}{L.~Baranyi},
\newblock \bibinfo{title}{Hydrodynamic and thermal study of a trapezoidal
  cylinder placed in shear-thinning and shear-thickening non-newtonian liquid
  flows},
\newblock \bibinfo{journal}{International Journal of Mechanical Sciences}
  \bibinfo{volume}{157} (\bibinfo{year}{2019}) \bibinfo{pages}{304--319}.
\bibitem[{Airiau and Bottaro(2020)}]{Airiau2020flo}
\bibinfo{author}{C.~Airiau}, \bibinfo{author}{A.~Bottaro},
\newblock \bibinfo{title}{Flow of shear-thinning fluids through porous media},
\newblock \bibinfo{journal}{Advances in Water Resources} \bibinfo{volume}{143}
  (\bibinfo{year}{2020}).
\bibitem[{Suresh et~al.(2018)Suresh, Kumar, Boro, Kumar, and
  Pugazhenthi}]{Suresh2016rhe}
\bibinfo{author}{K.~Suresh}, \bibinfo{author}{R.~V. Kumar},
  \bibinfo{author}{R.~Boro}, \bibinfo{author}{M.~Kumar},
  \bibinfo{author}{G.~Pugazhenthi},
\newblock \bibinfo{title}{Rheological behavior of polystyrene (ps)/co-al
  layered double hydroxide (ldh) blend solution obtained through solvent
  blending route: Influence of ldh loading and temperature},
\newblock in: \bibinfo{booktitle}{International Conference on Processing of
  Materials, Minerals and Energy (PMME)}, volume~\bibinfo{volume}{5},
  \bibinfo{year}{2018}, pp. \bibinfo{pages}{1359--1371}. \URLprefix \url{<Go to
  ISI>://WOS:000428718200178}. \DOIprefix\doi{10.1016/j.matpr.2017.11.222}.
\bibitem[{Liu et~al.(2018)Liu, Zhang, Zhu, Fu, Ma, and Li}]{Liu2018for}
\bibinfo{author}{C.~Liu}, \bibinfo{author}{Q.~Zhang}, \bibinfo{author}{C.~Zhu},
  \bibinfo{author}{T.~Fu}, \bibinfo{author}{Y.~Ma}, \bibinfo{author}{H.~Z. Li},
\newblock \bibinfo{title}{Formation of droplet and "string of sausages" for
  water-ionic liquid ( bmim pf6 ) two-phase flow in a flow-focusing device},
\newblock \bibinfo{journal}{Chemical Engineering and Processing-Process
  Intensification} \bibinfo{volume}{125} (\bibinfo{year}{2018})
  \bibinfo{pages}{8--17}.
\bibitem[{Du et~al.(2018)Du, Fu, Duan, Zhu, Ma, and Li}]{Du2018bre}
\bibinfo{author}{W.~Du}, \bibinfo{author}{T.~Fu}, \bibinfo{author}{Y.~Duan},
  \bibinfo{author}{C.~Zhu}, \bibinfo{author}{Y.~Ma}, \bibinfo{author}{H.~Z.
  Li},
\newblock \bibinfo{title}{Breakup dynamics for droplet formation in
  shear-thinning fluids in a flow-focusing device},
\newblock \bibinfo{journal}{Chemical Engineering Science} \bibinfo{volume}{176}
  (\bibinfo{year}{2018}) \bibinfo{pages}{66--76}.
\bibitem[{Sontti and Atta(2017)}]{Sontti2017cfd}
\bibinfo{author}{S.~G. Sontti}, \bibinfo{author}{A.~Atta},
\newblock \bibinfo{title}{Cfd analysis of microfluidic droplet formation in
  non-newtonian liquid},
\newblock \bibinfo{journal}{Chemical Engineering Journal} \bibinfo{volume}{330}
  (\bibinfo{year}{2017}) \bibinfo{pages}{245--261}.
\bibitem[{Chen et~al.(2020)Chen, Li, Song, Christopher, and Li}]{Chen2020mod}
\bibinfo{author}{Q.~Chen}, \bibinfo{author}{J.~Li}, \bibinfo{author}{Y.~Song},
  \bibinfo{author}{D.~M. Christopher}, \bibinfo{author}{X.~Li},
\newblock \bibinfo{title}{Modeling of newtonian droplet formation in power-law
  non-newtonian fluids in a flow-focusing device},
\newblock \bibinfo{journal}{Heat and Mass Transfer} \bibinfo{volume}{56}
  (\bibinfo{year}{2020}) \bibinfo{pages}{2711--2723}.
\bibitem[{Kokini et~al.(1984)Kokini, Bistany, and Mills}]{J1984pre}
\bibinfo{author}{J.~Kokini}, \bibinfo{author}{K.~L. Bistany},
  \bibinfo{author}{P.~Mills},
\newblock \bibinfo{title}{Predicting steady shear and dynamic viscoelastic
  properties of guar and carrageenan using the bird-carreau constitutive
  model},
\newblock \bibinfo{journal}{Journal of Food Science} \bibinfo{volume}{49}
  (\bibinfo{year}{1984}) \bibinfo{pages}{1569--1572}.

\end{thebibliography}

\end{document}